\documentclass[twocolumn,nofootinbib,amsmath,amssymb]{revtex4}

\usepackage{amssymb}
\usepackage{amsmath}
\usepackage{graphicx}
\usepackage{dcolumn}
\usepackage{bm}
\usepackage{color}
\usepackage{subfigure}

\newcommand{\be}{\begin{equation}}
\newcommand{\ee}{\end{equation}}

\begin{document}

\title{Cosmic Bandits: Exploration versus Exploitation in CMB B-Mode Experiments}

\title{Cosmic Bandits: Exploration versus Exploitation in CMB B-Mode Experiments}

\author{Ely D. Kovetz$^{1,2}$ and Marc Kamionkowski$^{2}$}

\affiliation{$^1$Theory Group, Department of Physics and Texas
     Cosmology Center, The University of Texas at Austin, TX
     78712, USA}
\affiliation{$^2$Department of Physics and Astronomy, Johns
     Hopkins University, Baltimore, MD 21218, USA} 
     
\begin{abstract}
A preferred method to detect the curl-component, or B-mode,
signature of inflationary gravitational waves (IGWs) in the
cosmic microwave background (CMB) polarization, in the absence of
foregrounds and lensing, is a prolonged integration over a single
patch of sky of a few square degrees. In practice, however,
foregrounds abound and the sensitivity to B modes can be
improved considerably by finding the region of sky cleanest of
foregrounds.  The best strategy to detect B modes thus involves
a tradeoff between exploration (to find lower-foreground
patches) and exploitation (through prolonged integration). This
problem is akin to the multi-armed bandit (MAB) problem in
probability theory, wherein a gambler faces a series of slot
machines with unknown winning odds and must develop a strategy
to maximize his/her winnings with some finite number of pulls.
While the optimal MAB strategy remains to be determined, a
number of algorithms have been developed in an effort to
maximize the winnings. Here, based on this resemblance, 
we tackle the search for IGW B modes with single frequency experiments
in the presence of spatially-varying 
foregrounds by developing adaptive survey strategies to optimize the
sensitivity to IGW B modes.  We demonstrate, using realistic
foreground models and taking lensing-induced B modes into account, that adaptive
experiments can substantially improve the upper bound on the
tensor-to-scalar ratio (by factors of 2--3 in single frequency experiments, 
and possibly even more). Similar techniques can be applied to other
surveys, including 21-cm measurements of signatures of the epoch
of reionization, searches for a stochastic primordial gravitational wave 
background, deep-field imaging by the James Webb Space
Telescope or various radio interferometers, and transient follow-up searches.
\end{abstract}

\maketitle

\section{Introduction}
Cosmology has become a science of surveys.  Ever larger surveys
are used to seek ever-more-subtle correlations to shed light on
novel early-Universe phenomena or the physics of galaxy
formation.  The separation of the signals of interest from
similar ones due to astrophysical foregrounds requires
more sensitive measurements and clever algorithms.  The issue of
foregrounds can also be dealt with by restricting the survey to
``clean'' regions, where the foregrounds are absent or at least
smaller in amplitude.  But finding these clean regions requires a
search which may then take time away from integration on a
single patch of sky.  Optimization of the sensitivity to a given
signal may thus involve a tradeoff between {\it exploration} of
several patches of sky, to find the cleanest one, and {\it
exploitation}, deep integration on a single patch.  What is the
best strategy, under these circumstances, to optimize the
sensitivity to the signal?  

This question is somewhat analogous to the multi-armed bandit (MAB) problem,
a well-known problem from probability theory and machine learning in 
computer science \cite{Robbins, Berry, PressTalk}.  In this problem, a 
gambler is faced with a set of slot machines with different reward
probability distributions and has to maximize the total reward
in a given number of plays, or actions. This is a classic
learning problem, as repeated plays allow the gambler to {\it
learn} the distributions of the different machines, with a
tradeoff between exploration and exploitation governed by the
total number of allowed plays. A popular manifestation of this
problem, which has garnered growing attention in recent decades,
is clinical trials \cite{Press}, where rewards---in the form of
survival/fatality---are of particular importance.
Theoretical study of the MAB problem has led to several theorems
regarding the ultimate prospects of solution methods in the
asymptotic limit of infinite number of plays
\cite{LaiRobbins}. In realistic scenarios, however, with only a finite
number of plays, one must resort to heuristic approaches, and
over the years several classes of these have been suggested in
the literature and compared empirically to some extent
\cite{Sutton,McGill,Gittins}.

In this paper we focus on the search for the curl, or B-mode,
signature of inflationary gravitational waves (IGWs) in the cosmic 
microwave background (CMB) polarization
\cite{Kamionkowski:1996zd,Seljak:1996gy,Kamionkowski:1999qc}.
These B modes are the target of a number of
ongoing and forthcoming CMB-polarization experiments
\cite{POLARBEAR,BICEP,QUIET,KECK,QUAD,SPTpol,ACTpol,ABS,EBEX,MAXIPOL,SPIDER,CLASS,CLOVER}
\footnote{In the past year, first detections of B-mode polarization from lensing of E-modes were announced
 \cite{Zaldarriaga:1998ar,SPTpolBModes, Ade:2014afa}, followed by a detection on degree scales \cite{Ade:2014xna}, 
whose source remains under dispute (whether it is primordial or due to foregrounds) \cite{Flauger:2014qra,Mortonson:2014bja, PlanckMay2014}.}. 
The strategy of many of these experiments is to integrate deeply
on a small patch of sky, as this optimizes the sensitivity to
IGW B modes in an experiment with fixed detector sensitivity, or
noise-equivalent temperature (NET), and duration
\cite{Jaffe:2000yt}.  Realistically,
though, these experiments will have to contend with foreground
emission from Galactic dust and synchrotron radiation
\cite{Verde:2005ff,Kogut:2007tq,Stivoli:2010rs,Fantaye:2011zq,O'Dea:2011kx,Clark:2012sa}.
Since the amplitudes of these foregrounds may vary considerably
from one region of the sky to another \cite{Fantaye:2011zq,O'Dea:2011kx,Clark:2012sa, PlanckSep2014},
the sensitivity
to IGWs may be improved considerably by integrating on the
cleanest patch.  While measurements (mostly unpolarized) at
other frequencies can be used to steer the experimentalist
toward a clean region of the sky \cite{KovKamExploration}, the {\it polarized}
foregrounds in the electromagnetic and spatial frequencies of
interest have only been measured to poor accuracy in the cleanest
regions of sky \cite{PlanckSep2014}.  One can thus do an initial
exploration of a broad region to find clean patches \cite{KovKamExploration}, 
but that then takes time away from exploiting any particular region.
An important challenge is thus to balance the tradeoff between
exploration and exploitation in an optimal way, given the limits
set by instrumental properties (including the total observation
time of the experiment) and the expected distribution of
foreground noise on the sky.

The purpose of this paper is to present a method inspired by
heuristic solutions to the MAB problem to optimally perform the
integration over sky patches so that noise from polarized
foregrounds is minimized and the strongest possible upper bound can be
placed on the amplitude of IGW B modes. We consider
several fiducial experiments with instrumental properties
representative of current and next-generation experiments, all
operating at a single frequency of $150\,{\rm GHz}$ (a value
common to many of the leading B-mode experiments) and focus on
the dominant foreground source at this frequency, which is
polarized emission from dust (PED) in the galaxy.

In order to forecast the variation of this foreground source
across the sky, we use the FGPol \cite{O'Dea:2011kx} foreground
templates for PED. We perform simulations of different survey
(bandit) strategies on patches of sky within a low-noise region
accessible from the South Pole, for which PED amplitudes are
randomly drawn from the FGPol template, and calculate the
improvement (or degradation) in the upper bound on the
tensor-to-scalar ratio $r$.  
While our analysis makes a number of
simplifications (although we {\it do} include lensing-induced B
modes \cite{Zaldarriaga:1998ar,SPTpolBModes}, an essential ingredient), our results
demonstrate that the adaptive survey strategies we consider
provide considerable advantage over prolonged integration on
naively-chosen patches.

While our focus here is on CMB polarization, the methods
described in this work can also be applied to other
observations in cosmology and astrophysics, such as 21-cm
measurements \cite{Furlanetto, Morales, Pritchard}, searches for a 
stochastic primordial gravitational wave background \cite{PrimGWBack1,PrimGWBack2}, 
deep-field telescope imaging \cite{Windhorst:2005as,Stiavelli}, and transient
searches \cite{Djorgovski:2011rv}. We discuss potential issues
pertaining to such applications, but leave their full study to
future work.
 
The plan of the paper is as follows:  In Section II we describe
the PED templates used in our analysis, discuss the instrumental
noise of our fiducial experiments and present the statistical
tools for estimating the errors in measurements of the relevant
power spectra. In Section III, we describe how we construct 
and test adaptive survey strategies based on machine-learning heuristics, 
and explain our
prescription for simulating adaptive B-mode experiments. We
present our results in Section IV and discuss several
assumptions and possible additional implementations in Section
V. We conclude in Section VI.

\section{PED Foreground}


In order to remove the different foreground contributions, most
experiments operate at several frequencies and use component
separation \cite{Stivoli:2010rs,Fantaye:2011zq} or
template-based techniques \cite{Fantaye:2012ha} to extract as
clean a signal as possible. In any such process, residuals
remain at some level and will hinder the ability to detect the
desired signal.

The major contributions of polarized foreground noise in the
relevant frequency range ($\sim20-300\,{\rm GHz}$) of CMB
experiments are sourced by PED in the Galaxy
and by synchrotron radiation. Synchrotron is more dominant at
lower frequencies ($\lesssim100\,{\rm GHz}$), while PED
overwhelms the CMB at higher frequencies ($\gtrsim100\,{\rm
GHz}$). The well-known CMB foreground ``sweet spot'' is around
$90\,{\rm GHz}$, where the noise sources are similar in
amplitude and both are comparably low.

For simplicity, we shall address fiducial experiments operating
at a single frequency of $150\,{\rm GHz}$, which is adopted by
many of the sub-orbital polarization experiments (see \cite{KovKamExploration}
for a complementary discussion of multi-frequency approaches). Therefore, PED
would be the major source for concern in terms of foregrounds
and its subtraction would be difficult. To estimate the sky
variation of the PED power spectrum, we use the FGPOL templates
\cite{O'Dea:2011kx,Clark:2012sa}\footnote{http://www3.imperial.ac.uk/people/c.contaldi/fgpol}. These
templates are based on a three-dimensional bi-symmetric spiral
\cite{O'Dea:2011kx} model of the Galactic magnetic field
(including the turbulent component) and are normalized according
to the results of WMAP \cite{Kogut:2007tq} so that the average
dust polarization fraction outside the WMAP P06 polarization
mask \cite{Page:2006hz} is $3.6\%$.
Prior to the release of polarization results from the Planck experiment, 
the best constraint on
the dust polarization fraction at higher frequencies came from
the partial sky ($f_{\rm sky}=17\%$) measurement at $353\,{\rm GHz}$ of
the ARCHEOPS balloon-borne experiment \cite{Benoit:2003hg},
which detected a polarization fraction around $4-5\%$ in the
Galactic plane and a maximum of $10-20\%$ in some localized
clouds. According to the more recent Planck results based on 
$353\,{\rm GHz}$ measurements \cite{PlanckSep2014}, the average 
polarization fraction in high-galactic latitudes is around $10\%$,
while there is evidence, albeit with high uncertainty, that there
exist regions of sky with considerably lower PED amplitudes. 
We shall therefore consider $3.6\%$ and $10\%$
normalizations as conservative and worst-case scenarios,
respectively.

\begin{figure}
\includegraphics[width=0.48\linewidth]{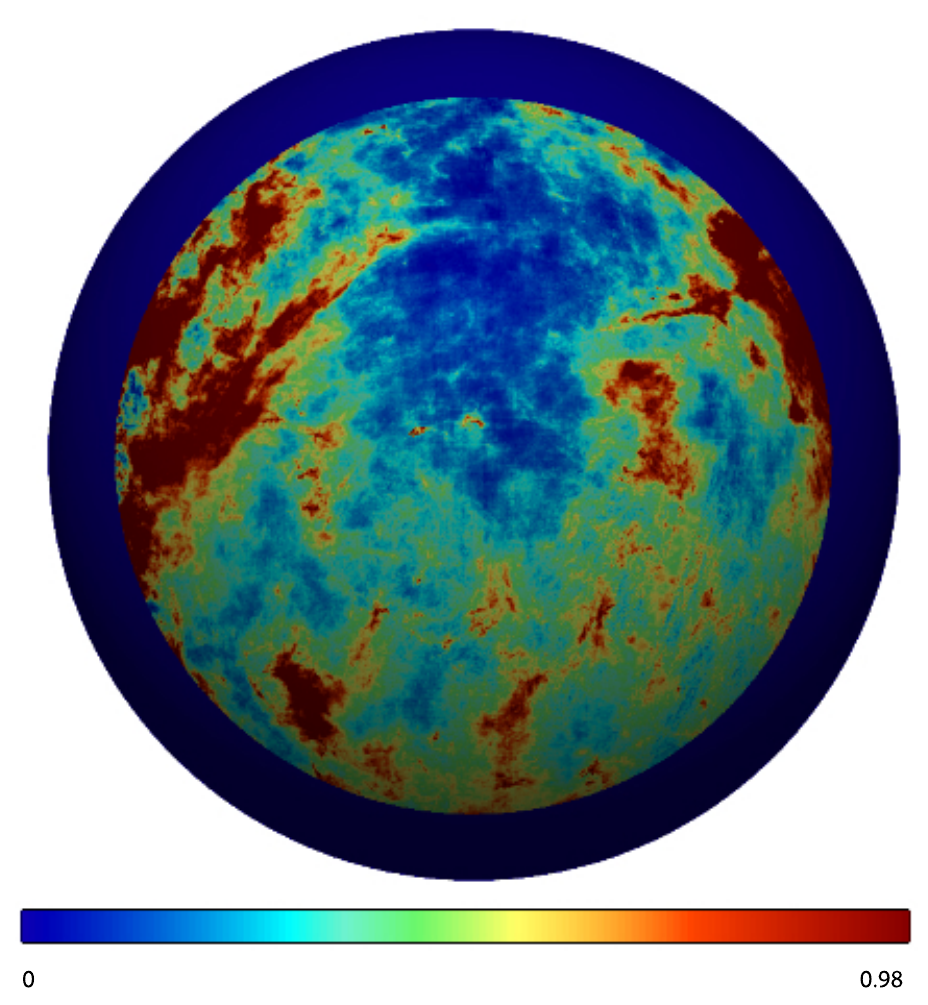}
\includegraphics[width=0.46\linewidth]{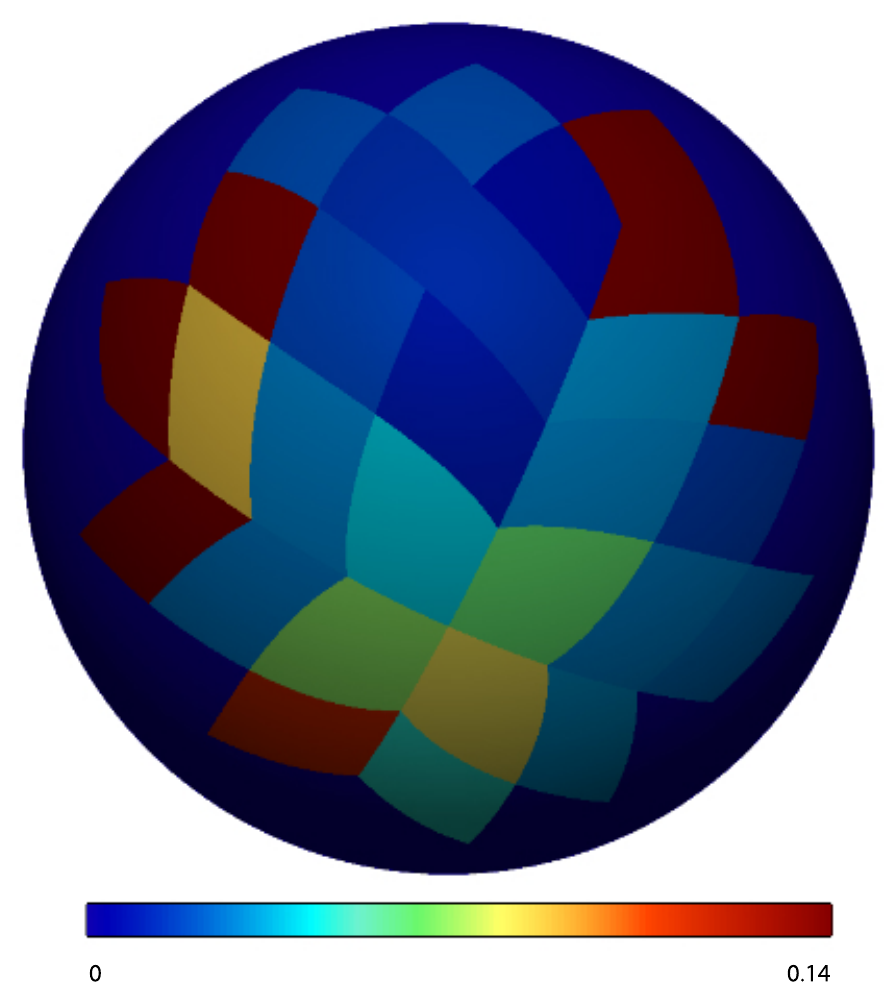}
\caption{{\it Left:} The polarization amplitude
     $\sqrt{Q^2+U^2}$ in a $45^{\circ}$-radius region centered
     around $(b,l)=(-69.5, 241.5)$, taken from the FGPol
     template \cite{O'Dea:2011kx} (in units of $\mu{\rm K}$). {\it Right:} The average
     of the $Q$ and $U$ variances ($[\mu{\rm K}^2]$) in non-overlapping
     $15^{\circ}\times15^{\circ}$ patches in the same region
     (roughly matching the HEALPIX resolution Nside$\,=\!\!4$ used
     for this plot).}
\label{fig:PatchSkyVariance}
\end{figure}

In Fig.~\ref{fig:PatchSkyVariance} we plot the high-resolution
FGPol template for a $45^{\circ}$-radius low-noise region
centered around $(b,l)=(-69.5, 241.5)$, which is accessible by
ground-based experiments such as BICEP, POLARBEAR, the
KECK Array, QUAD, QUIET and SPTPOL. We also plot the average 
of the variance in the $Q$ and $U$ polarization components in
non-overlapping patches of $15^{\circ}\times15^{\circ}$ inside
this region.

In order to estimate the PED amplitude in a given patch, we
follow the prescription of Ref.~\cite{Clark:2012sa}. The PED
angular power spectrum is assumed to obey a power law,
\be
\label{eq:cldust}
     \frac{\ell(\ell+1)}{2\pi}C^D_\ell = A\ell^m,
\ee
where the power-law index $m$ is fixed to its full-sky best-fit
value $m=-0.22$ in this template, and $A$ is allowed to vary
between different patches.  To estimate $A_p$ for a given patch
$p$, we calculate the variances $\sigma_Q^2 = \langle Q^2
\rangle_p - \langle Q\rangle_p^2$ and $\sigma_U^2 = \langle U^2
\rangle_p - \langle U\rangle_p^2$ of the polarization in both
the $Q$ and $U$ components of the patch, and infer a
power-spectrum amplitude from each, using the relation
\begin{equation}
     \sigma^2 = \frac{1}{4\pi}\sum_{\ell=2}^{\ell_{\rm max}}
     (2\ell+1)C^D_{\ell}B^2_{\ell}(\theta_s).
\label{eq:sigmainversion}
\end{equation}
We then take the average of the resulting amplitudes to be the
PED B-mode power-spectrum amplitude in the desired patch, under
the assumption that power is equally distributed between the E
and B modes\footnote{This assumption is not supported by recent
Planck measurements, which suggest that $C_l^{BB}\sim0.5C_l^{EE}$, 
but this difference can be absorbed in the large uncertainty regarding
the average polarization fraction---which we already take into 
account---and therefore has no critical influence on our results.}.

\begin{figure}
\includegraphics[width=\linewidth]{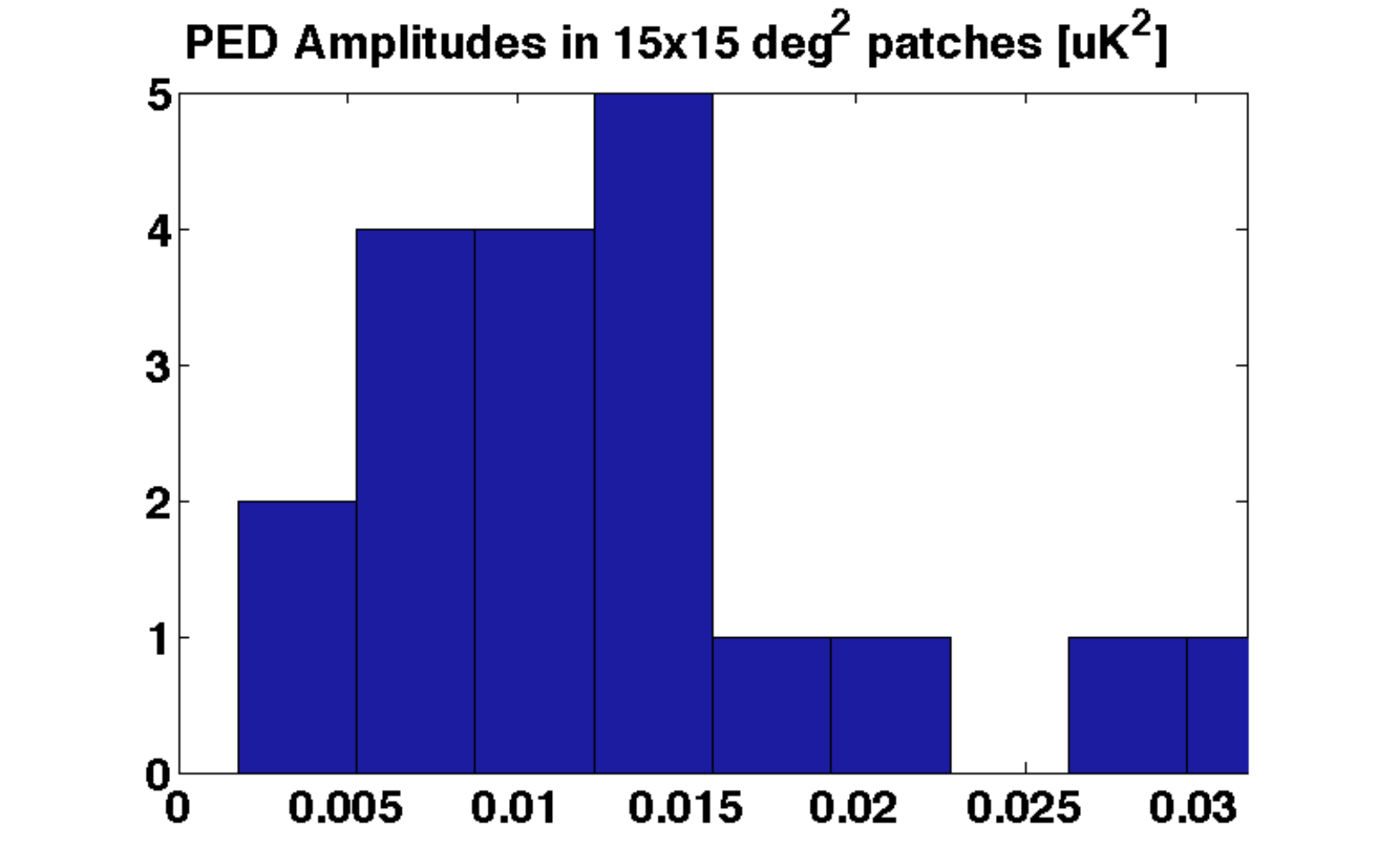}
\caption{A histogram of dust amplitudes $A_p$ in $19$ patches of
     $15^{\circ}\times15^{\circ}$ within the region plotted in
     Fig.~\ref{fig:PatchSkyVariance}, calculated using
     Eq.~(\protect\ref{eq:sigmainversion}), after a cutoff at the
     $67$th-percentile to remove the noisiest patches from our
     sample altogether.}
\label{fig:PatchHistVariance}
\end{figure}

Finally, to remain with a realistic set of possible patches, we
immediately apply a cutoff to remove the  noisiest patches from
our sample. The choice of cutoff represents prior knowledge
regarding the expected PED amplitudes in this region given the
templates and results of previous surveys. We choose a cutoff of $33\%$
throughout.  A histogram of PED amplitudes $A_p$ in the
remaining $19$ patches within the region shown in
Fig.~\ref{fig:PatchSkyVariance} is plotted in
Fig.~\ref{fig:PatchHistVariance}, with a normalization of
$3.6\%$. We see that the amplitudes in this sample still vary
over more than an order of magnitude.

Before moving on to survey optimization methods, we wish to examine
the expected signal, noise and foreground amplitudes.
We defer the instrumental noise calculation (as well as the statistical estimators 
used to measure CMB power spectra) to the Appendix, and present the results here.

We consider three types of
fiducial experiments whose parameters are given in
Table~\ref{table:exppars}. Experiment 1 is inspired by the
POLARBEAR experiment \cite{POLARBEAR} and observes a single
$15^{\circ}\times15^{\circ}$ patch. Experiment 2 has similar
properties, but a smaller sky coverage, optimized for detecting
the peak of the primordial B-mode power spectrum at
$\ell\sim10^2$ (without de-lensing, the optimal size may be a
bit larger \cite{Kesden:2002ku}). Experiment 3 is a fiducial
lower-cost, lower-resolution experiment targeted at the larger scale
primordial B-mode signal (hence the larger sky coverage). We
assume an observing efficiency of $20\%$ for all three
experiments.

\begin{table}[h!]
\begin{center}
\begin{tabular}{|c|c|c|c|c|c|}
\hline 
Experiment  &   $\theta_{\rm fwhm}$  & $f_{\rm sky}$ & $T$ & $s$=NET$_{\rm array} $\\
 & [arcmin]  &  [\%]&  [years] & [$\mu{\rm K}\sqrt{\rm sec}$] \\\hline 
1  & 3.5 & 0.55 & 2  & $\frac{480\sqrt{2}}{\sqrt{1274}}=19$\\\hline
2 & 5  & 0.014 & 2 & 15 \\\hline
3 & 30  & 1.52 & 6 & 25 \\\hline
\end{tabular}
\caption{Parameters for three CMB polarization experiments we use for our analysis. Here, $s$ is the single-detector polarization sensitivity divided by the square root of the number of detectors.}
\label{table:exppars}
\end{center}
\end{table}

\begin{figure}[t!]
\includegraphics[width=\linewidth]{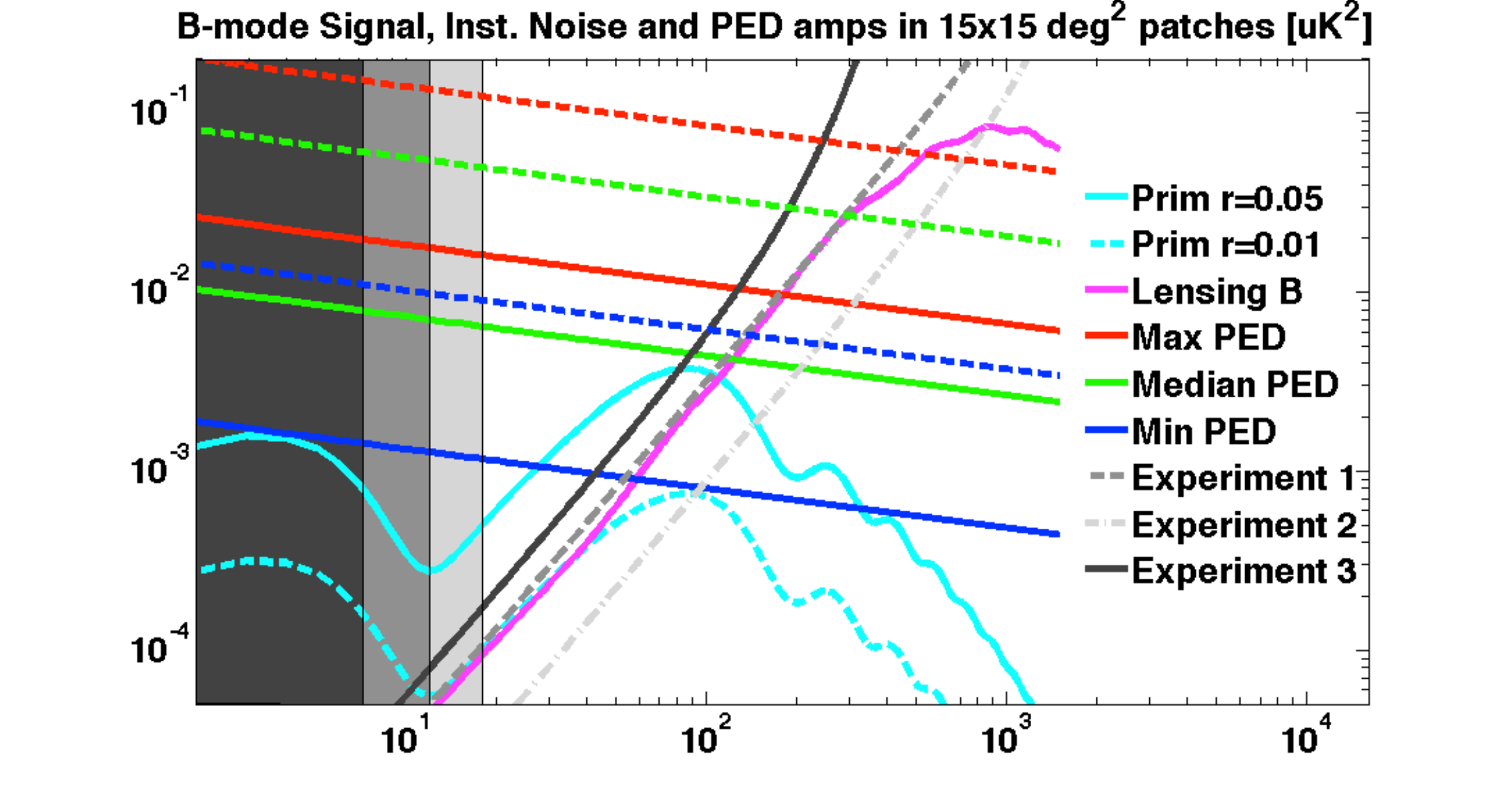}
\caption{A plot of the relevant power spectra: the primordial
     B modes for $r=0.05$ and $r=0.01$ (in solid and dashed
     cyan), the lensing-induced B-mode contribution (magenta), the
     instrumental noise for our three fiducial experiments (gray
     dashed, dot-dashed and solid lines)---each limited to a
     different $\ell$ range (shaded gray regions)---and the maximum,
     median, and minimum PED power spectra in $19$ non-overlapping
     $15^{\circ}\times15^{\circ}$ patches, for $3.6\%$ (solid)
     and $10\%$ (dashed) normalization values (in blue, red and
     green).}
\label{fig:SignalAndNoise}
\end{figure}

In Fig.~\ref{fig:SignalAndNoise}, we plot the expected noise
power spectrum of our fiducial experiments, together with the
primordial B-mode power spectrum for tensor-to-scalar ratios
$r=0.05,\,0.01$. The lensing contribution to the B-mode signal is
plotted separately. We also plot the PED power spectra for the
maximum-, median-, and minimum-amplitude patches in our sample, for two fiducial
normalization levels, corresponding to the values $3.6\%$ and
$10\%$ of average dust polarization fraction outside the
Galactic plane.

We can see that the primordial contribution to the signal peaks
once at very low $\ell$ (due to the reionization contribution) and
then again around $\ell\sim10^2$, while B modes from lensing
become dominant at smaller scales and peak around
$\ell\sim10^3$.  At low $\ell$, the signal lies below most
fiducial noise levels, except perhaps the optimal patches of sky
under the assumption that the average dust polarization fraction
is not too high. Therefore, it is clear that foregrounds pose a
serious challenge to a single-frequency measurement, and a
systematic method to identify lower-noise patches may be
useful.

\section{Exploration vs. Exploitation}

We now describe the methodology for constructing adaptive survey
strategies to identify the optimal patches within an initial set of candidates.
Using consecutive short-time measurements of B modes over the patches, 
we will show how heuristic machine-learning algorithms can be used 
to converge onto the optimal patch.

\subsection{Constructing and Evaluating Adaptive Strategies}

A strategy to find the best patch uses a set of estimates for
the expected rewards of each patch in order to decide which one to
select at each step.  The expected (or mean) reward of each patch
$patch$ is called its action-value, and we denote it by
$\mu^{*}(p)$.
We define $V_i$, the {\it reward} of the each of these short measurements, 
as $V_i=-\widehat{A_p}_i$, where $\widehat{A_p}_i$ is the estimated dust-amplitude in 
patch $p$ from measurement $i$. The goal of the survey strategy will thus be
to identify the patch with the highest true reward $\mu^{*}(p)$

A natural way to estimate the action-value of a patch after $t$
steps is through sample-averaging of its previous rewards. If at
step $t$ a patch $p$ has been chosen $N_t(p)$ times and has
yielded a set of rewards $V_1,V_2,\dots,V_{N_t(p)}$, then its
action-value estimate will be given by
\be
\mu_t(p) = \frac{\mu_0(p)+V_1+V_2+\dots+V_{N_t(p)}}{N_t(p)+1},
\ee
where the values $\mu_0(p)$ are defined by some chosen method of initialization.
The law of large numbers then guarantees that $\mu_t(p)\to \mu^{*}(p)$ as $N_t(p)\to\infty$. 

The standard figure of merit for the success of a proposed
algorithm to solve a problem of this type is its total regret. if
$V^{*}=\underset{p}{\operatorname{max}}\,\mu^{*}(p)$ is the
expected reward of the optimal patch $p^{*}$, then after $t$ plays
the total regret is given by
\be
     R_t=\left\langle\sum\limits_{t=1}^{T}\left[
     V^{*}-\mu_t(p)\right]\right\rangle=\sum\limits_{p}\left\langle
     N_t(p)\right\rangle\Delta_p,
\label{eq:regret}
\ee
where $\Delta_p=V^{*}-\mu^{*}(p)$ is the gap between the optimal
reward and the expected reward of patch $p$. Therefore, a good
strategy ensures smaller counts for larger gaps.

The performance of any strategy depends on the similarity
between the optimal patch and the other patches. The hardest problems
will have similar-looking patches with slightly different means. 
In the context of MAB solutions, Lai and Robbins \cite{LaiRobbins} have 
derived a lower bound for the asymptotic total regret in the limit of unlimited 
number of plays. As we demonstrate below, this bound applies quite well 
to the algorithms we consider below for B mode survey strategies as well. 
According to their theorem,
\be
     \lim_{t\to\infty}R_t\geq
     \log{t}\sum\limits_{p|\Delta_p>0}
     \frac{\Delta_p}{KL(\mathcal{P}(p)||\mathcal{P}(p^{*}))},
\ee
where $KL(\mathcal{P}(p)||\mathcal{P}(p^{*}))$ is the
Kullback-Leibler (KL) distance \cite{KLdiv} between the probability
distributions of a patch $p$ and the optimal patch $p^{*}$. If the
rewards have normal distributions,
$\mathcal{N}(\mu^{*}(p),\sigma_p)$ and
$\mathcal{N}(\mu^{*}(p^{*}),\sigma_{p^{*}})$, the KL distance
between two patches is given by \cite{Kullback}
\be
     KL(\mathcal{P}(p)||\mathcal{P}(p^{*})) =
     \ln{\frac{\sigma_{p^{*}}}{\sigma_p}}+
     \frac{\sigma^2_{p}+\left(\mu^{*}(p^{*})-\mu^{*}(p)\right)^2}
     {2\sigma_{p^{*}}^2}-\frac{1}{2}
\ee

In practice, achieving logarithmic total regret is a challenge,
requiring an ideal tradeoff between exploration and
exploitation. It is easy to see from Eq.~(\ref{eq:regret}) that
an algorithm that never explores will have linear total regret
and the same is true for an algorithm that explores forever.

\subsection{Families of Heuristic Solution Methods}

We briefly describe a set of algorithms that represent some of
the prevailing heuristics in the literature to solve problems of this type.

\begin{itemize}

\item{Greedy}

The most naive approach to this optimization problem is simply to select
at each observation ``step'' $t$ the patch with the highest estimated action-value
\be
     p_t=\underset{p}{\operatorname{argmax}}\, \mu_t(p).
\ee 
This method merely exploits past knowledge in order to maximize
the immediate reward (hence {\it greedy} \cite{Cormen}) and
results in linear total regret. In the absence of any initial
knowledge regarding the expected rewards (setting all
$\mu_0(p)=-M$ where M is very large), this method chooses one of the
patches randomly on the first step and then continuously exploits
it, without any further exploration. 

\item{$\epsilon$-greedy}

A simple enhancement to the regular greedy method is to
occasionally force the selection of a patch at random, regardless
of the action-value estimates. In the $\epsilon$-greedy method, 
at probability $1-\epsilon$ the greedy patch is chosen, while at
probability $\epsilon$ a patch $p$ (out of a total $n_p$ patches)
is chosen uniformly at random.
In the limit $\epsilon\to 0$, this reduces to the regular greedy
method. When $\epsilon=1$, this method forever exploits all patches
uniformly. 

\item{Decaying $\epsilon$-greedy}

While perpetual exploration ensures the convergence of all the
action-value estimators $\mu_t(p)\to \mu^{*}(p)$ (since
$N_t(p)\to\infty$ for all patches), it also means that exploration 
continues even after the optimal patch is identified with great (and ever-growing)
confidence.
To solve this problem, a further enhancement to this method is
to decrease the exploration probability $\epsilon$ with time. In practice, however, 
it is hard to choose a single decay strategy for
$\epsilon$ that works well under different circumstances
(different reward distributions).

\item{Optimistic initialization}

We implement the absence of prior knowledge by taking the limit
$\mu_0(p)\to-\infty$ (we take $\mu_0(p)=-M$ where M is very large).  
Any other choice for the initial
action-value estimates would introduce a bias in all the methods
described so far, but an effective choice of this bias can be
used as a means of forcing an early stage of increased
exploration. For example, by choosing optimistically-high
initial values in the greedy method, we delay the convergence
onto a single patch until the initial values no longer dominate
the sample average. Until that happens, some knowledge will have
been gathered about all patches, ensuring a more informed greedy
choice from then on, at the expense of this initial stage of
exploration.

\item{Probability matching (Boltzmann)}

An obvious path towards more sophisticated algorithms is to vary
the probability with which inferior patches are explored, instead
of sampling them uniformly as in $\epsilon$-greedy. After all,
$\epsilon$-greedy is likely to keep exploring extremely
unfavorable patches even after they are known to be vastly inferior
to others.  The straightforward solution is to set the
probability to explore each patch according to its estimated
value. In the Boltzmann method (or Thompson sampling
\cite{Thompson}), for example, the probability to choose patch $p$
at play $t$ is given by
\be
     \text{Prob}_t(p)=\frac{e^{\mu_t(p)/\tau}}{\sum\limits_{p'}e^{\mu_t(p')/\tau}},
\ee
where $\tau$ is a positive parameter, often referred to as the
temperature. In the high-temperature limit, this reduces to the
$\epsilon=1$ uniform method, while in the $\tau\to 0$ limit we
retrieve the regular greedy method with absence of initial
knowledge.

Although some quantitative studies have been performed to
compare the Boltzmann and $\epsilon$-greedy methods, no
consistent conclusions as to which is superior have been reached
\cite{Sutton,McGill}. Both methods depend on the tuning of their
respective free parameter and one performs better than the other
under different circumstances.
We shall use the Boltzmann method with optimistic
initialization, as our experience deems this favorable.

\item{Upper confidence bound}

The upper confidence bound (UCB) method \cite{Auer} tackles the
exploration-versus-exploitation tradeoff according to the
principle that the more uncertain we are about the action-value 
of a patch, the more important it is to explore it further
(since it could turn out to be the optimal one).
In this method, we estimate an upper confidence $U_t(p)$ for
each action value such that $\mu(p) \leq \mu_t(p) + U_t(p)$ with
high probability, and then select the action that maximizes the sum $\mu_t(p)+U_t(p)$
\be
     p_t =\underset{p}{\operatorname{argmax}} \left\{\mu_t(p)+U_t(p)\right\} .
\ee
The confidence bound depends on the number of times $p$ has been
selected (as increased selection reduces the uncertainty
bound).  When the rewards are Gaussian, a useful upper bound is
given by $c\,\sigma_p/\sqrt{N_t(p)}$, where $c>0$ is some
constant. As a prior for the upper bound is used in this method, it is
often referred to as the ``UCB Bayesian'' method.

\end{itemize}

\begin{figure*}
\includegraphics[width=0.8\linewidth]{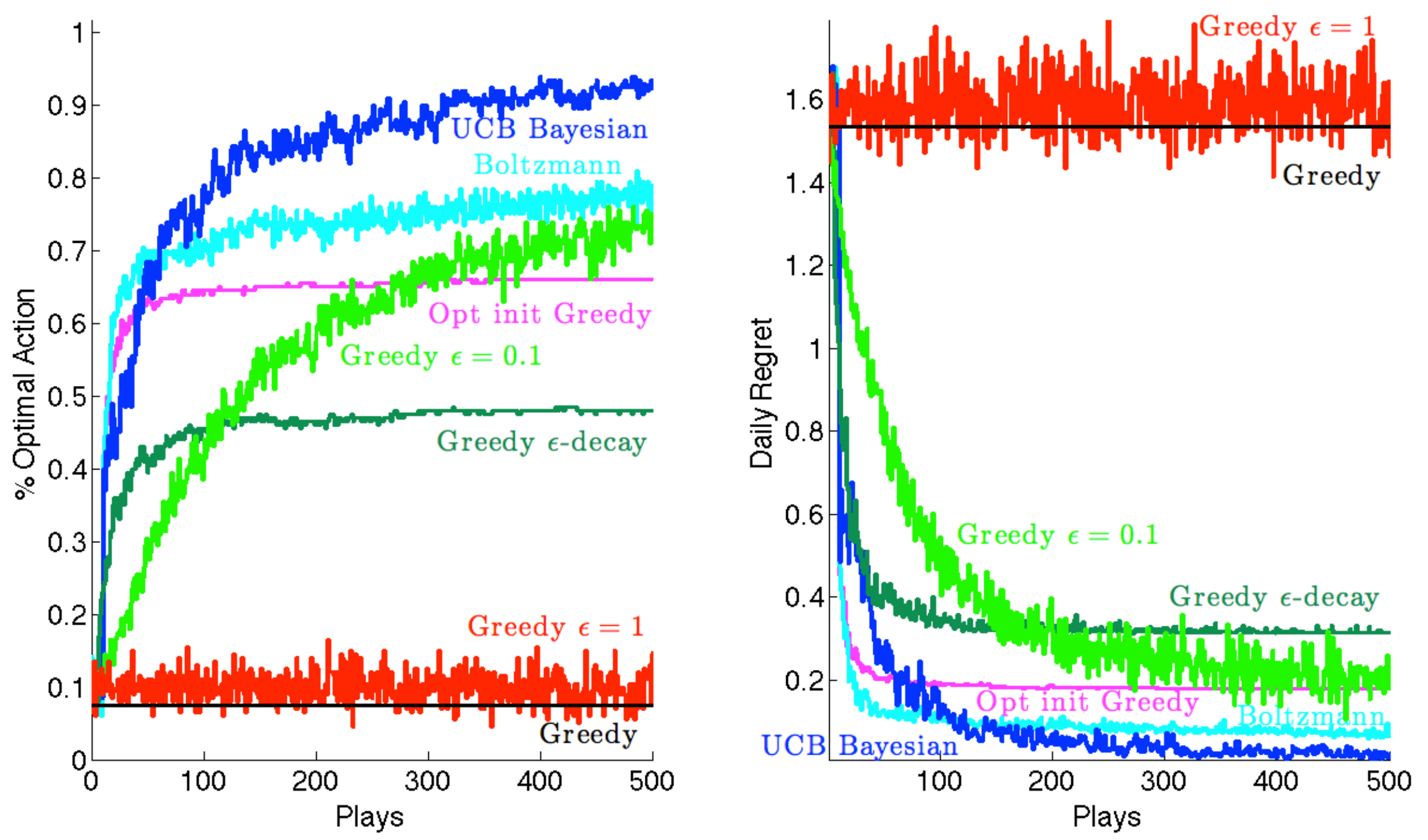}
\caption{Different algorithms balance exploration and
     exploitation differently. {\it Left:} The percentage of
     simulations in which the optimal patch was chosen in each
     play. The UCB method reaches a higher percentage than all
     other methods as the number of plays is increased. Greedy
     and $\epsilon$-greedy with $\epsilon=1$ choose the optimal patch
     $10\%$ of the time, which is expected with a total of $10$
     patches. {\it Right:} The average regret at each play. Notice
     that the Boltzmann method is inferior to UCB in the long
     run, but it reduces the daily regret faster than UCB at the
     beginning. }
\label{fig:SimpleAlgsDaily}
\end{figure*}

\begin{figure}[h!]
\includegraphics[width=0.49\linewidth]{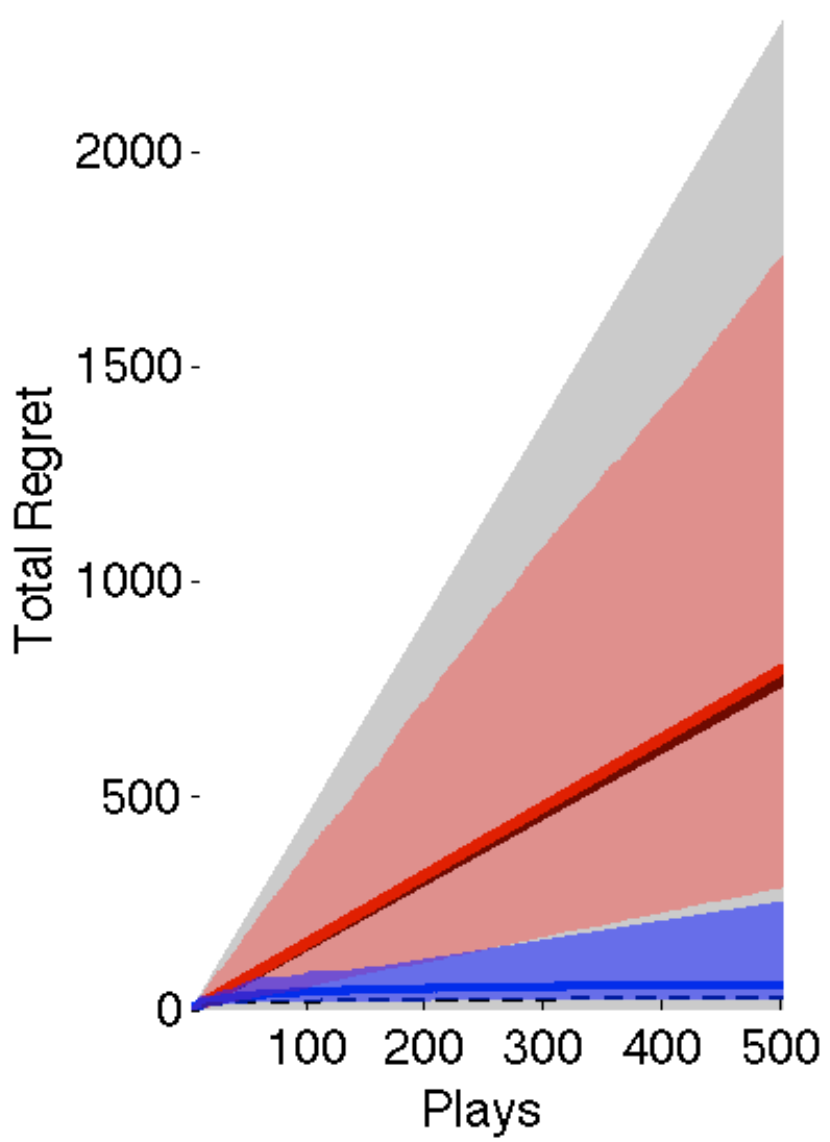}
\includegraphics[width=0.49\linewidth]{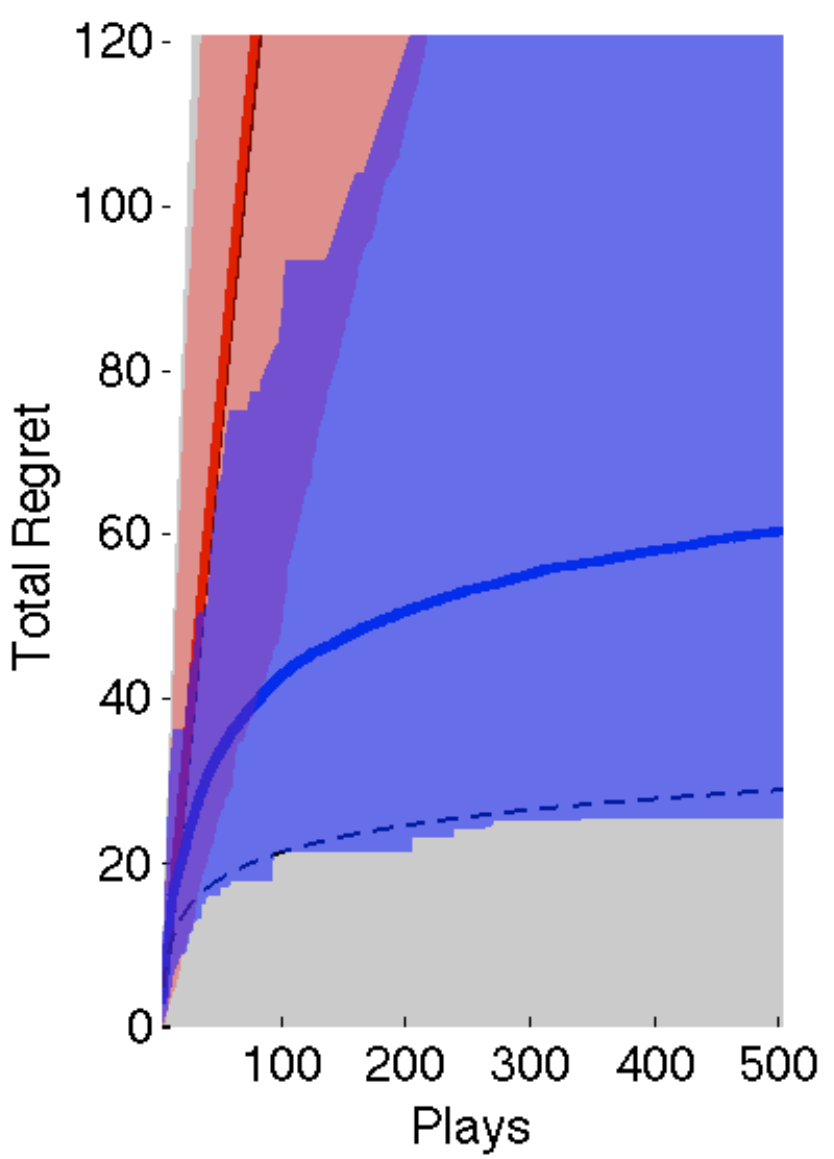}
\caption{{\it Left:} The average and ranges of total regret for greedy (range: shaded gray, average: solid black), $\epsilon$-greedy with $\epsilon=1$ (red) and UCB (blue). {\it Right:} Zooming in $\rightarrow$ the UCB method assumes logarithmic behavior, similar to that of the Lai and Robbins asymptotic bound (in dashed-black), after a relatively small number of plays.}
\label{fig:SimpleAlgsTotal}
\end{figure}

Before moving on to the implementation of these heuristic
strategies in B-mode experiments, we demonstrate their
performance for a simple problem with $10$ patches with reward
distributions $\mathcal{N}(\mu^{*}_p,1)$, where the mean rewards
$\mu^{*}_p$ are themselves drawn from a normal distribution,
$\mathcal{N}(0,1)$.  In Fig.~\ref{fig:SimpleAlgsDaily} we plot
the behavior of each of the methods with time. Most methods
converge onto the optimal patch in greater percentages and
decrease their daily regret as time goes by. The UCB method
seems more successful eventually than others, while the
second-best Boltzmann method is the most efficient after a very
limited number of plays.  In Fig.~\ref{fig:SimpleAlgsTotal} we
compare the total regret for three algorithms: greedy, $\epsilon$-greedy
with $\epsilon=1$, and UCB. We plot the average in solid lines
and the full range of performance in the corresponding shaded
regions. The greedy method will always have the widest range of
performance, as it chooses and stays with the optimal patch or
worst patch at $10\%$ probability. The UCB method has logarithmic
total regret, much like the Lai and Robbins asymptotic bound,
also plotted.

\subsection{Testing a B-mode adaptive strategy}

We now apply the heuristic adaptive-survey strategies 
discussed above to simulated CMB-polarization maps.  As detailed
below, each ``step'' will be a short-duration
measurement of the polarization on one of $n_p$ patches of sky
available to observe.  It is assumed that the data from this
short-duration experiment is analyzed under the null hypothesis
of no gravitational waves ($r=0$) to obtain an estimate (albeit
very rough) of the dust amplitude in that patch.  The ``reward''
from that ``pull'' is then the dust amplitude inferred from that
measurement multiplied by $-1$ (so that subsequent measurements
are steered toward {\it lower} dust amplitudes).  The chosen 
algorithm then uses whatever information it has collected from
measurements done so far on dust amplitudes to decide which
patch to observe next.  At no point does the algorithm make use
of any prior information about $A_p$, nor does it make any use
of the error estimates in Eq.~(\ref{eqn:sigmaprone}) for $r$.

In more detail, to test the adaptive strategies, we first simulate
an observed sky as follows:  We first apply an upper cutoff in
amplitude to the desired-size patches in the South Pole region
FGPol template to remain with a certain percentage of
lowest-noise patches, representing our prior knowledge regarding
the target area, either from the noise templates themselves or
from other surveys.  As mentioned above, for all experiments
we limit our sample to patches with PED amplitudes below the
$67$th percentile. We then randomly draw a subset of $n_p$ dust amplitudes from the truncated sample. 
These are drawn at the onset of each simulation and then kept fixed.

Meanwhile, we choose the time-step size, $t_{\rm step}$, typically a few observation
days. We also use the fiducial experimental parameters in
Table~\ref{table:exppars} to set the parameters $s$, $T$,
$f_{\rm sky}$, and $\sigma_b$.  We then use
Eq.~(\ref{eq:estimatorerror}) to calculate the ``1-sigma'' error
$\sigma^{\widehat{A}}$ with which the chosen experiment can 
measure the dust amplitude of a patch in a single time step.  (This
$\sigma^{\widehat{A}}$ corresponds to the width
of the likelihood functions in Fig.~\ref{fig:likelihoods}).

Finally, we apply the strategies to the simulated sky to
allocate the amount of time spent on each patch.  At each time
step $i$, we do the following:

\begin{itemize}

\item Choose which patch to observe according to the strategy we employ.

\item Retrieve a reward $V$ for this patch for this time
step through a random draw from a normal distribution,
$V= -{\rm max}(0,\mathcal{N}(A_p,\sigma^{\widehat{A}}))$.  The
minus sign is chosen so that the methods converge onto
lower-noise patches.

\item Update our action-value estimate $\mu_t(p)$ of the
chosen patch $p$ according to the reward.

\item Decide which patch to observe at the next time step
according to the strategy or stop the simulation if the total
observation time has been exhausted.  This decision is based
solely on the averages we have, at this step, of draws from the
$\widehat{A_p}$ maximum likelihood distribution.

\item Calculate the regret, measured by the difference in the
noise amplitude between the chosen patch and the lowest-noise
patch and update the total regret.  We also calculate the overall
figure-of-merit (the smallest detectable tensor-to-scalar ratio
at ``1-sigma'') at each time $T^i$ by plugging in the total
observation times spent so far on each patch $t^i_p$ (where
$\sum\limits_{p=1}^{n_p}t^i_p=T^i$) and the (known) dust
amplitudes $A_p$ into Eq.~(\ref{eq:r}). The regret and the overall
figure-of-merit are calculated but {\it never} used during the experiment. 
They are calculated only to allow us to evaluate at the end of the experiment
how well the strategy has converged to the cleanest patch of sky 
and how well it has improved the smallest B-mode amplitude $r$ that 
can be distinguished from a null result.

\end{itemize}

Our choice of parameters for the different methods:

\begin{itemize}
\item In the absence of initial knowledge: $\mu_0(p)=-3\sigma^{\widehat{A}}$
\item For $\epsilon$-greedy, we choose $\epsilon=0.1$.
\item Our $\epsilon$-decay strategy is $\epsilon={\rm min}(1,n_p\sigma^{\widehat{A}}/\sqrt{t})$.
\item For Optimistic Initialization: $\mu_0(p)=3\sigma^{\widehat{A}}$.
\item For the Boltzmann method, we choose $\tau=0.001$.
\item For UCB we take $c=1/2$:
$U_t(p)=\sigma^{\widehat{A}}/2\sqrt{N_t(p)}$.
\end{itemize}

We then repeat the full list of steps above $1,000$ times 
in order to acquire an ensemble of simulations for
comparison. We thus explore the improvement in the upper
bound on the tensor amplitude $r$ obtained with one strategy to
that obtained with another strategy. What changes between the
simulations is the set of dust amplitudes $A_p$ that we draw
each time from the FGPol template, as well as the random choices made by some of the algorithms.
In the next Section we study the performance of each of the strategies above for our
fiducial experiments (see  Table~\ref{table:exppars}) and for
the best-, conservative- and worst-case scenarios regarding the
PED normalization. 

\section{Results}

\begin{figure*}[th!]
\includegraphics[width=0.66\linewidth]{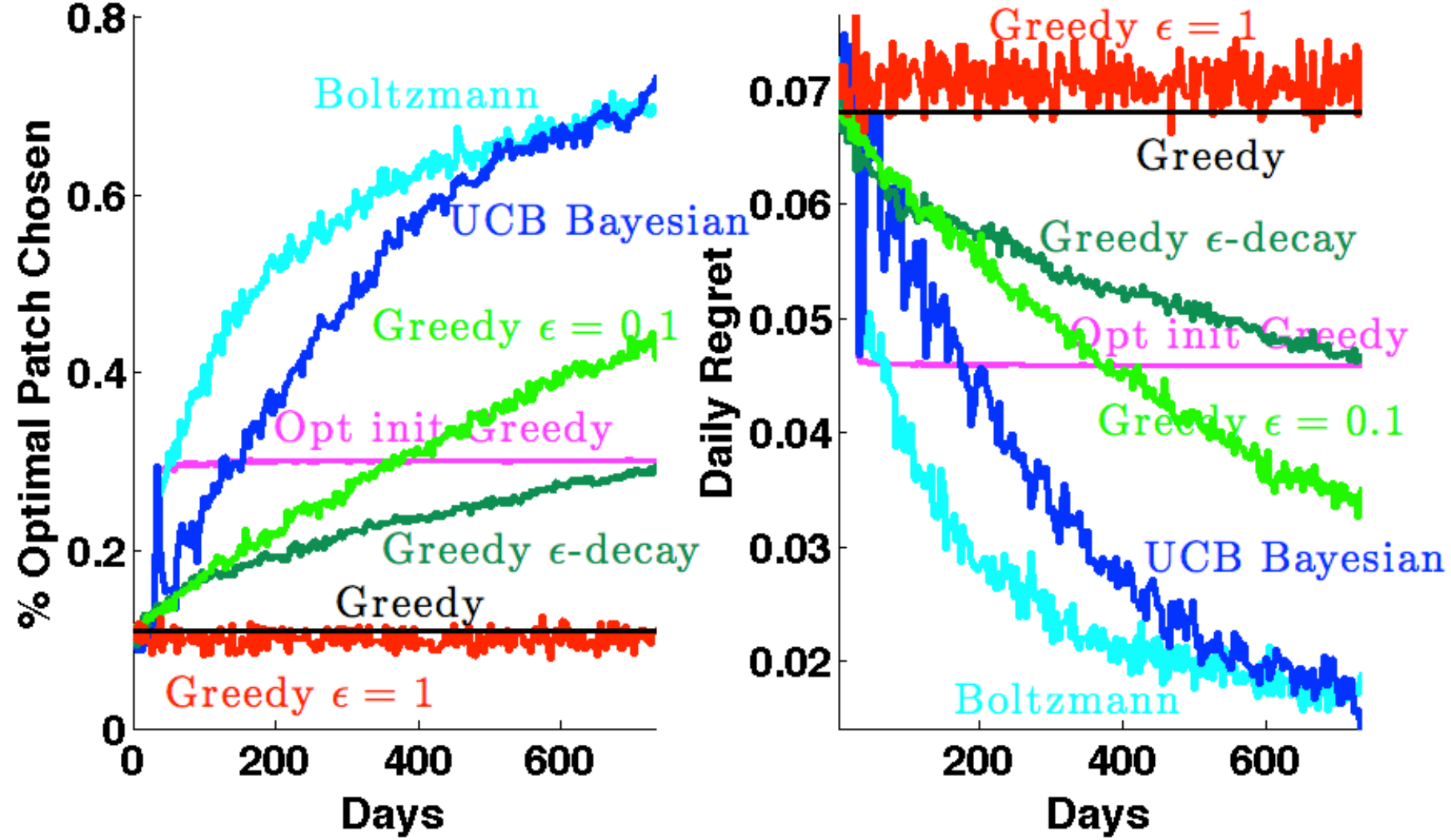}
\includegraphics[width=0.32\linewidth]{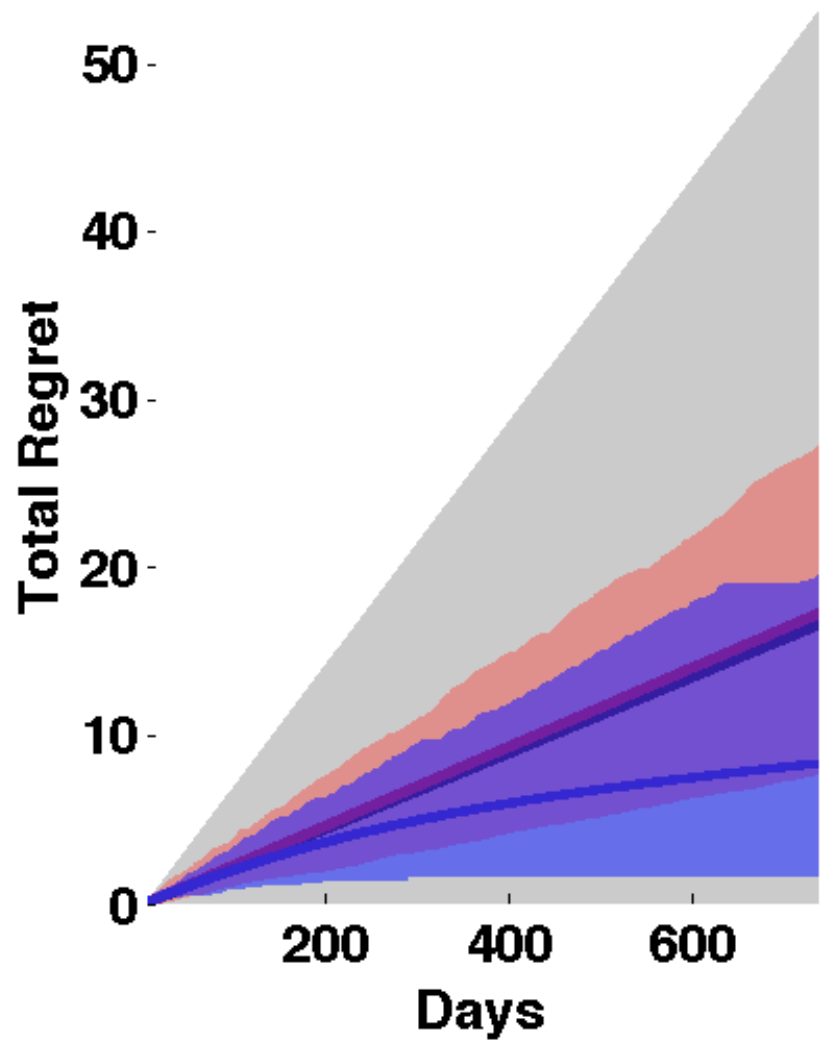}
\includegraphics[width=0.66\linewidth]{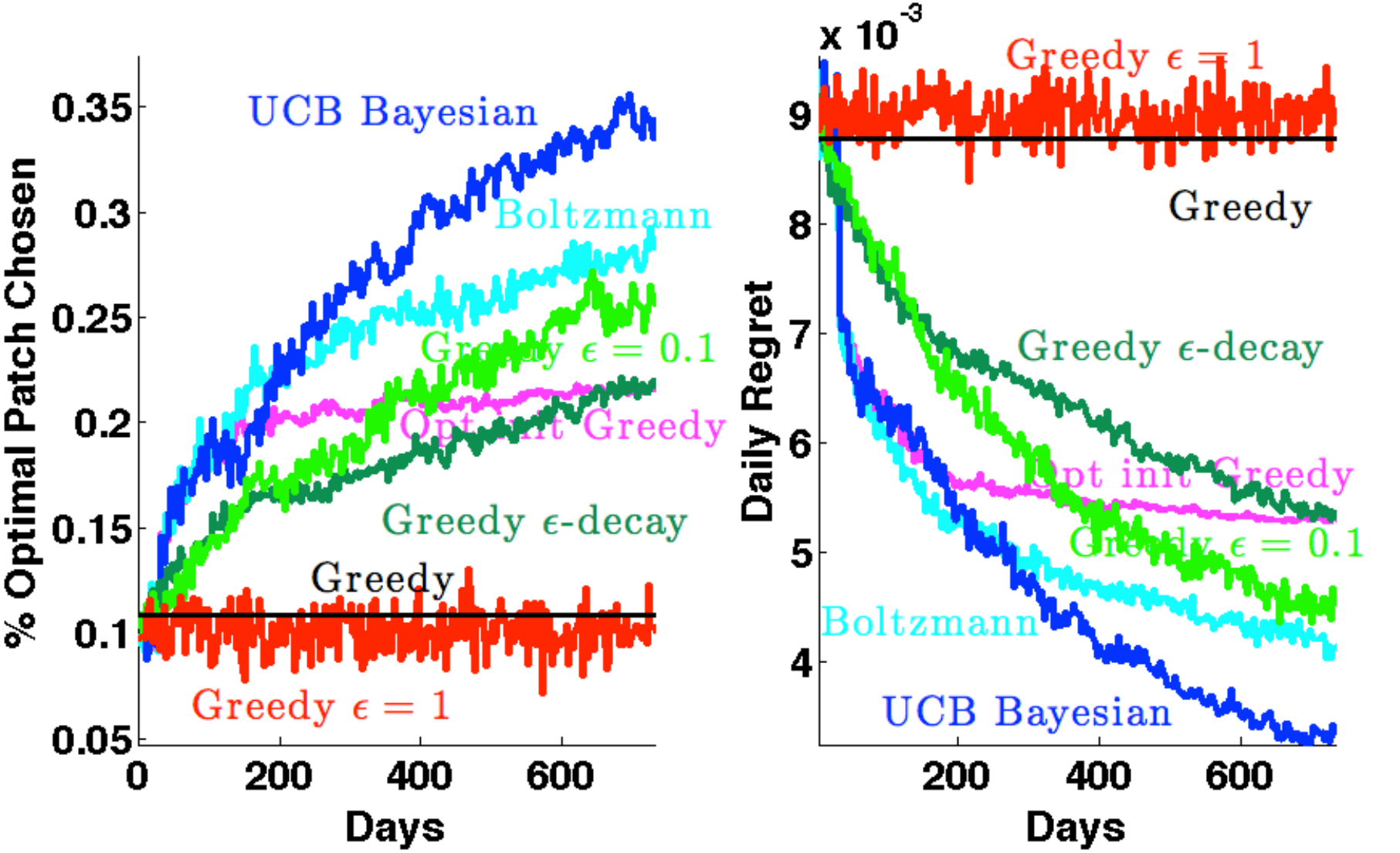}
\includegraphics[width=0.33\linewidth]{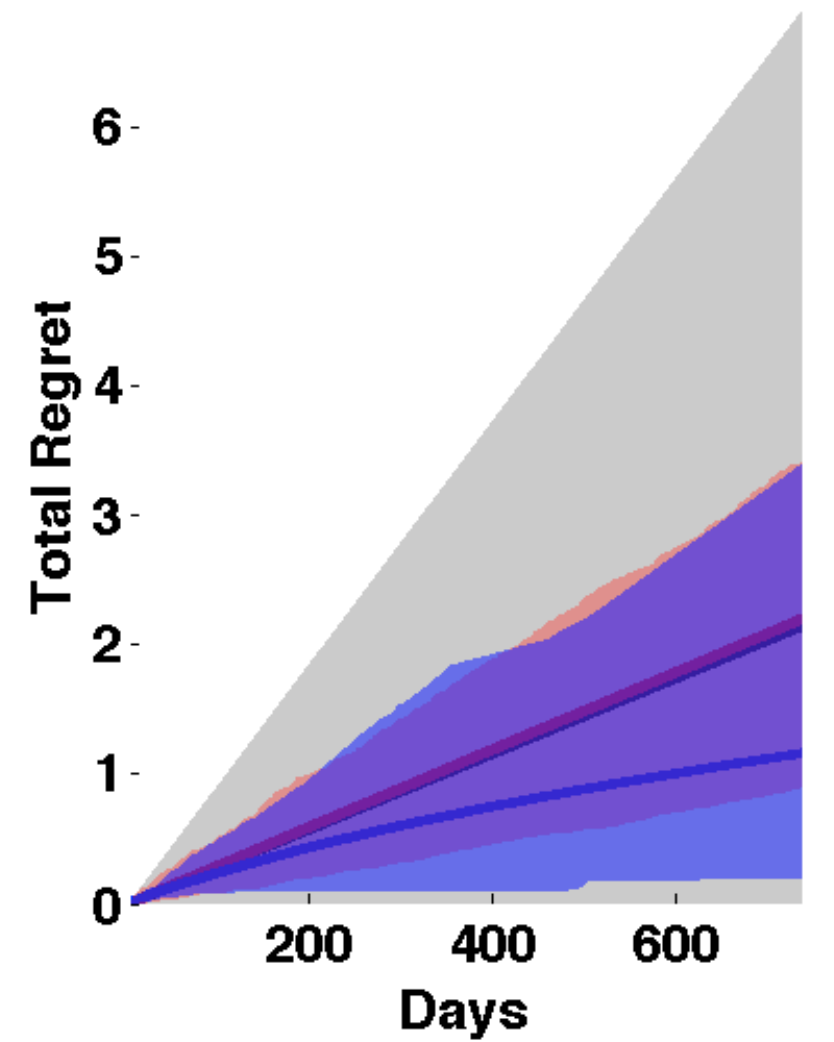}
\caption{The percentage of $1000$ simulations in which the
     optimal patch was chosen per day ({\it left}), the average
     daily regret ({\it center}) and the average total regret
     ({\it right}) for Experiment 1. {\it Top Row}: A pessimistic scenario of 
     no-delensing and $10\%$ average dust polarization fraction. 
     {\it Bottom Row}: The optimistic case of $3.6\%$ normalization 
     and $80\%$ de-lensing. We see that the UCB method ultimately chooses
     the optimal patch in larger percentages compared to other methods (although the Boltzmann method ramps up its performance more quickly at the onset). The UCB method achieves the lowest total regret on average,
     and even in its worst case (top of shaded-blue region), it
     is relatively close to the averages of greedy (black line, almost overlapping the red line, with gray shading) and $\epsilon$-greedy with $\epsilon=1$ (red), whose own  worst cases are roughly 4 and 10 times the UCB average. 
} 
\label{fig:ExampleExp}
\end{figure*}

We are now ready to test our strategies on simulations of
B-mode experiments. 
We first address the extreme cases (see Fig.~\ref{fig:likelihoods}) of $10\%$ average dust 
polarization fraction with no-delensing (a pessimistic scenario) and 
$3.6\%$ with an efficient de-lensing process, leaving only a $20\%$
residual (an optimistic scenario). Focusing on Experiment 1, which has properties similar
to the POLARBEAR experiment \cite{POLARBEAR}, we compare the performance of the
different strategies in Fig.~\ref{fig:ExampleExp} using our
simulations. We see that the UCB method fares better than other
methods in identifying the optimal patch, although it does not achieve the same rate of success
as in our simple Gaussian test case above. With a distribution
of foreground amplitudes taken from the FGPol templates in the
region of Fig.~\ref{fig:PatchSkyVariance}, after a cutoff at the
$67$th percentile, the UCB method manages to converge onto  the
optimal patch after 2 years of observation in only $\sim80\%$ of the simulations
in the pessimistic scenario and in less than half of the simulations in the 
optimistic one. Nevertheless, we see that the total regret in the optimal methods 
(Boltzmann, UCB) is much lower than with naive methods. The total
regret in the greedy method, which is the default version of a
POLARBEAR-type experiment, is roughly 2--3 times higher when comparing
the average or worst-case performance to those of UCB. Greedy
with $\epsilon=1$, which corresponds to an SPT-type experiment,
where a larger region is uniformly observed (or several patches
uniformly integrated over), is better in the worst case than
greedy, but also never reaches below the average of the UCB method in the pessimistic scenario
(and barely below the average of UCB in the optimistic scenario), even in its best-case performance.

\begin{figure*}[th!]
\includegraphics[width=0.32\textwidth]{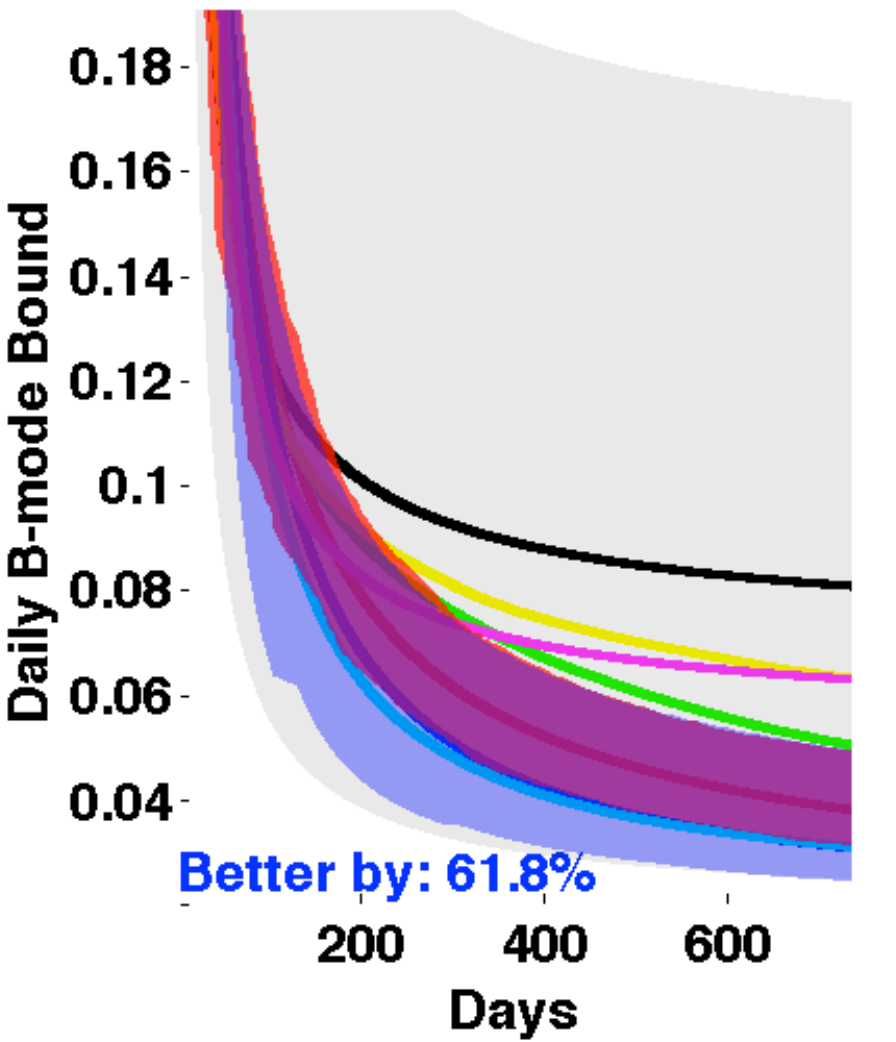}
\includegraphics[width=0.32\textwidth]{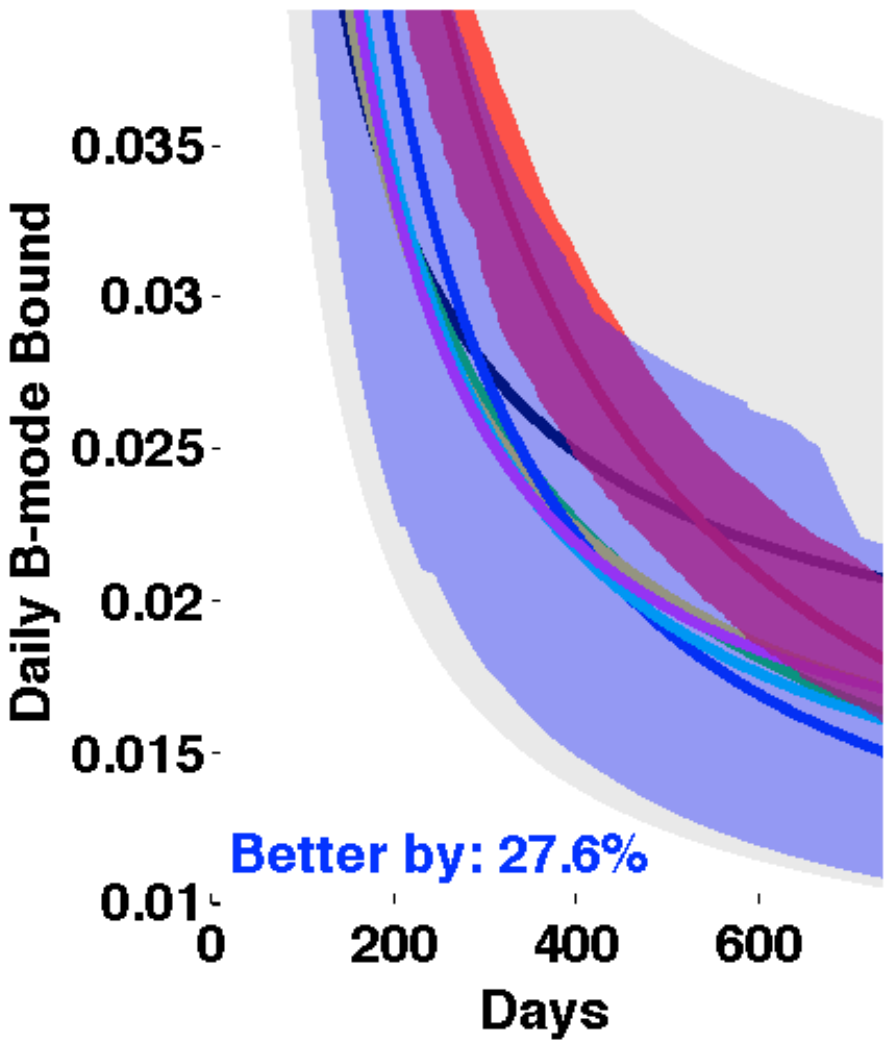}
\includegraphics[width=0.32\textwidth]{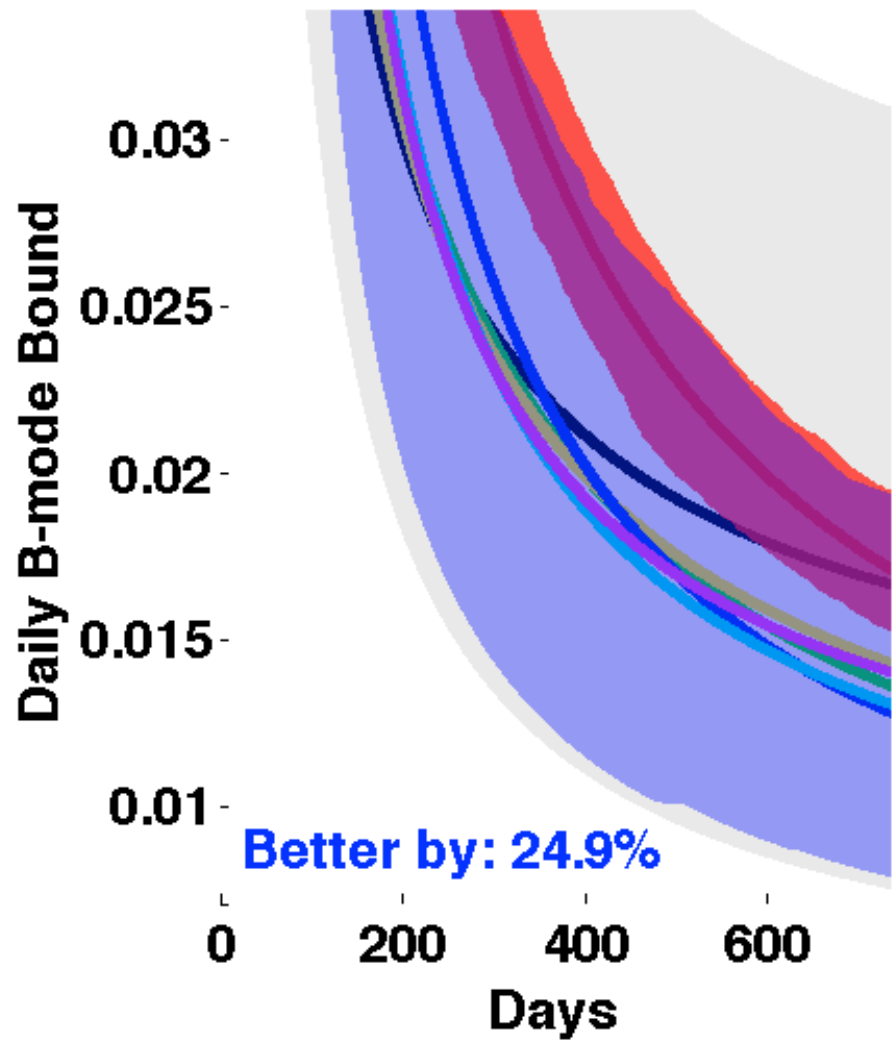}
\includegraphics[width=0.32\textwidth]{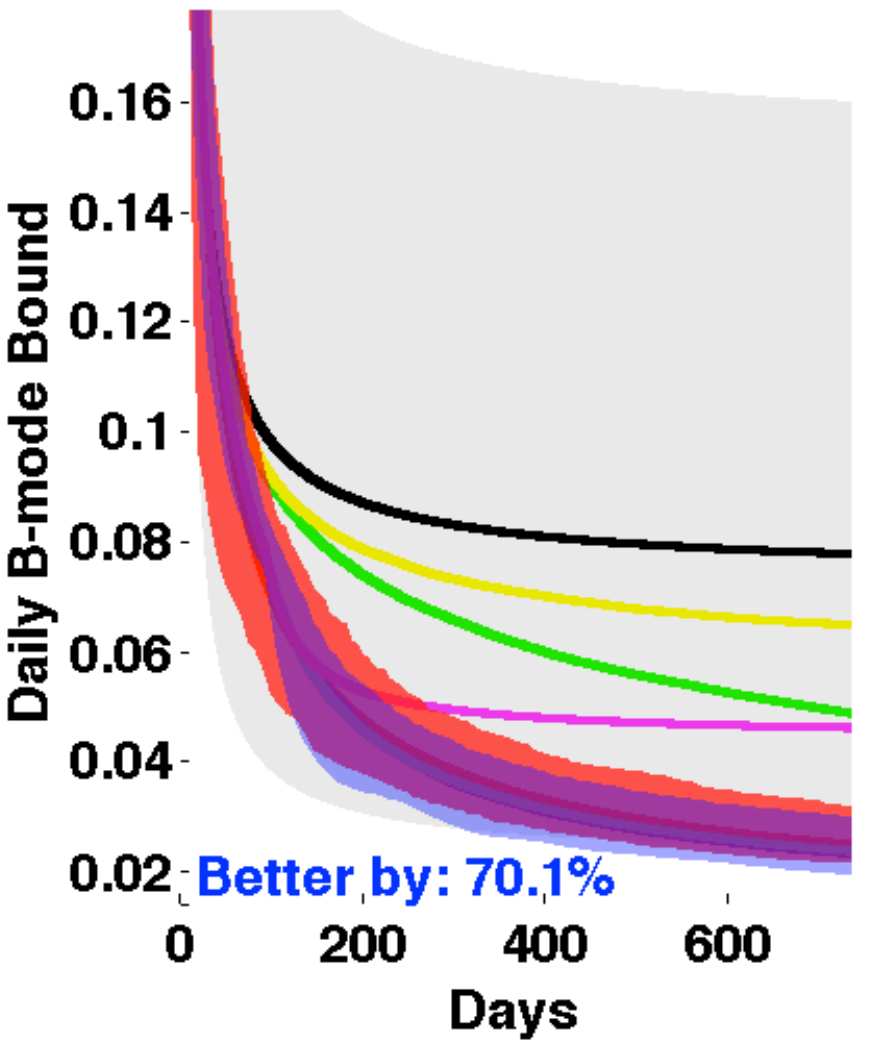}
\includegraphics[width=0.32\textwidth]{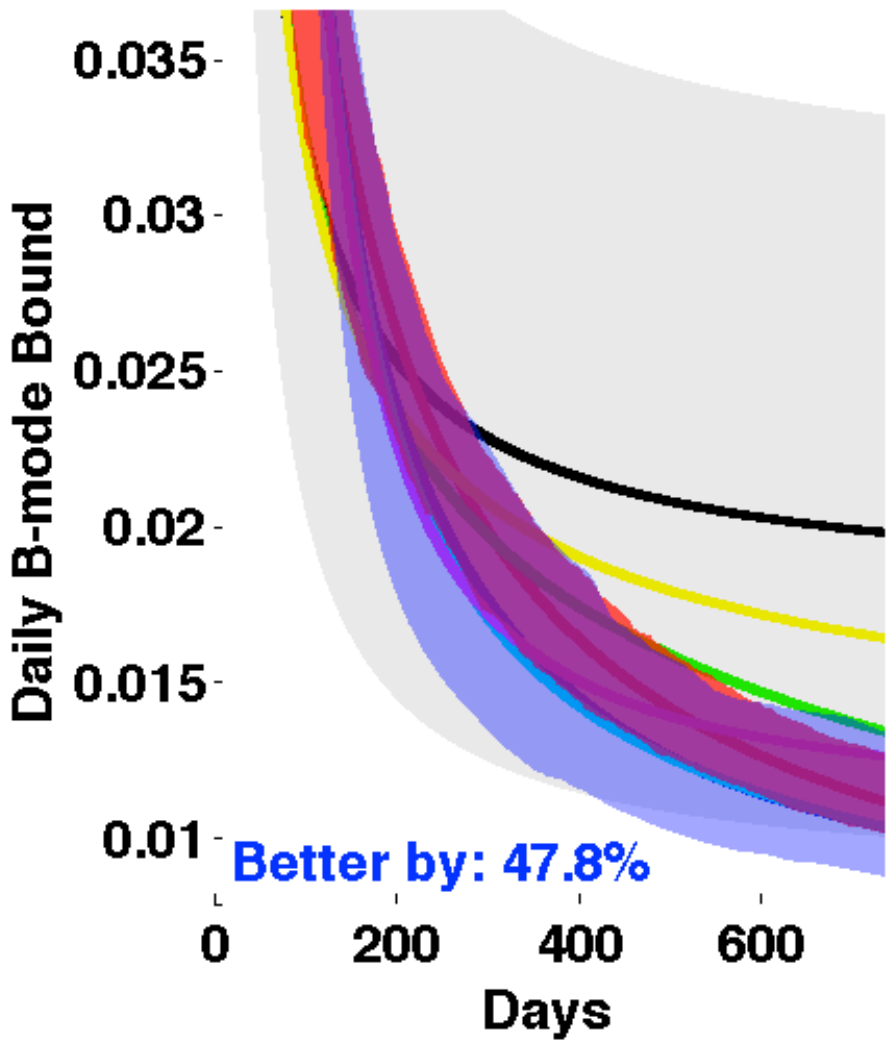}
\includegraphics[width=0.32\textwidth]{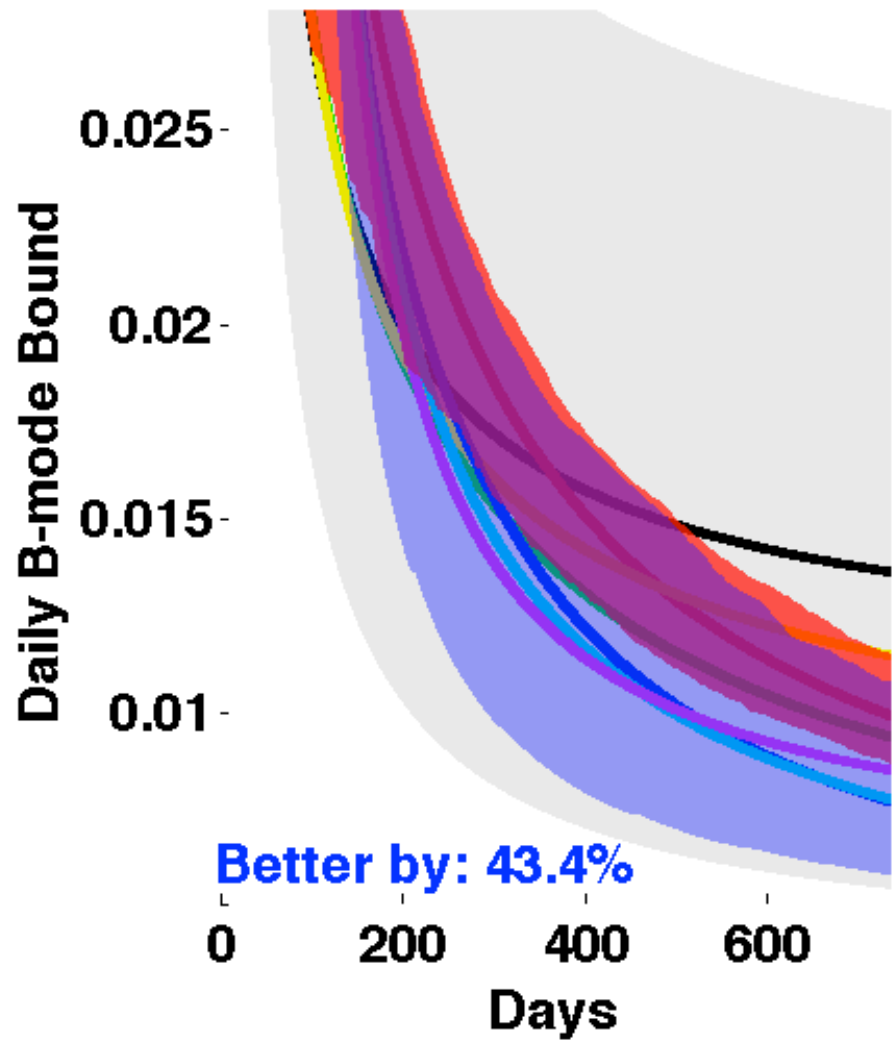}
\includegraphics[width=0.32\textwidth]{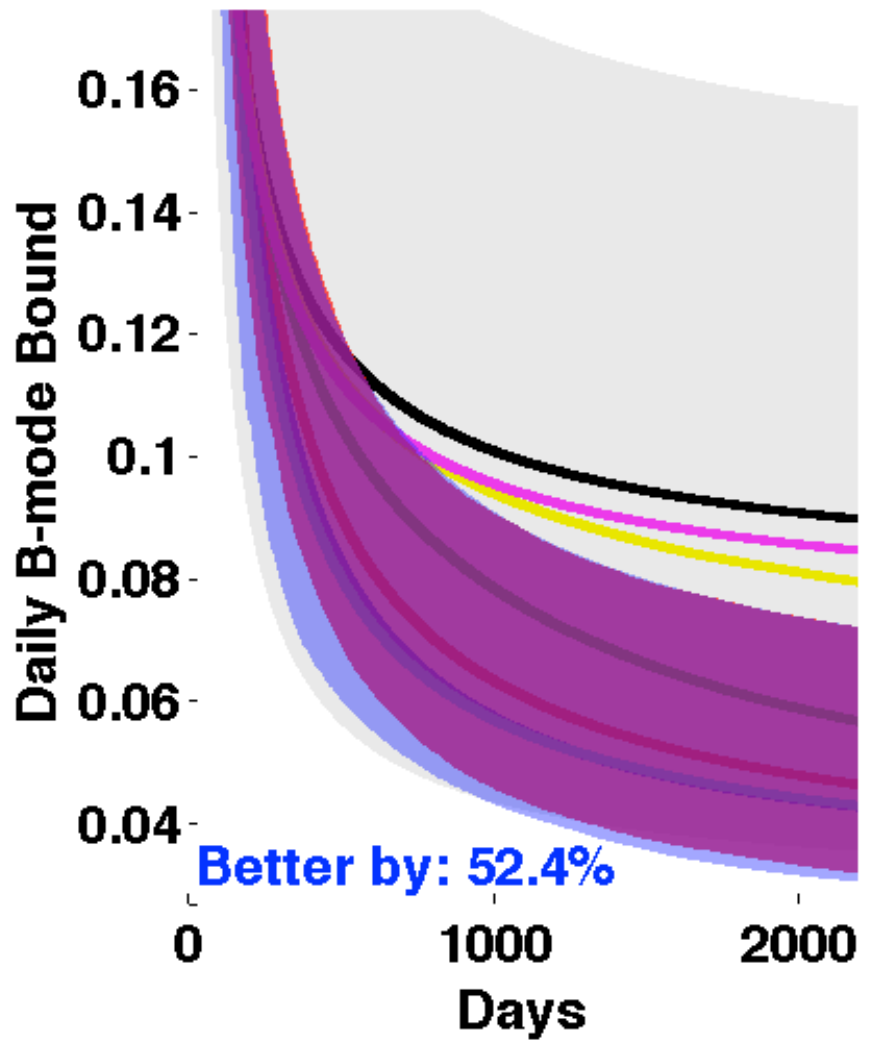}
\includegraphics[width=0.32\textwidth]{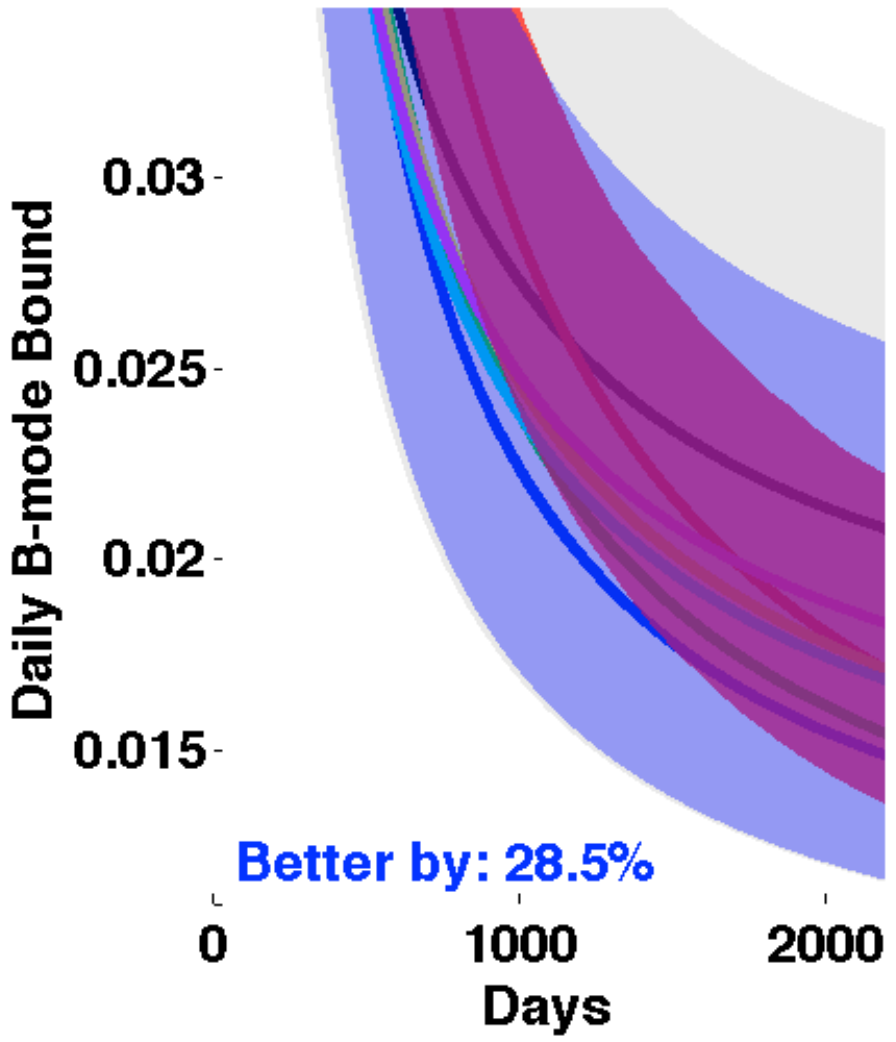}
\includegraphics[width=0.32\textwidth]{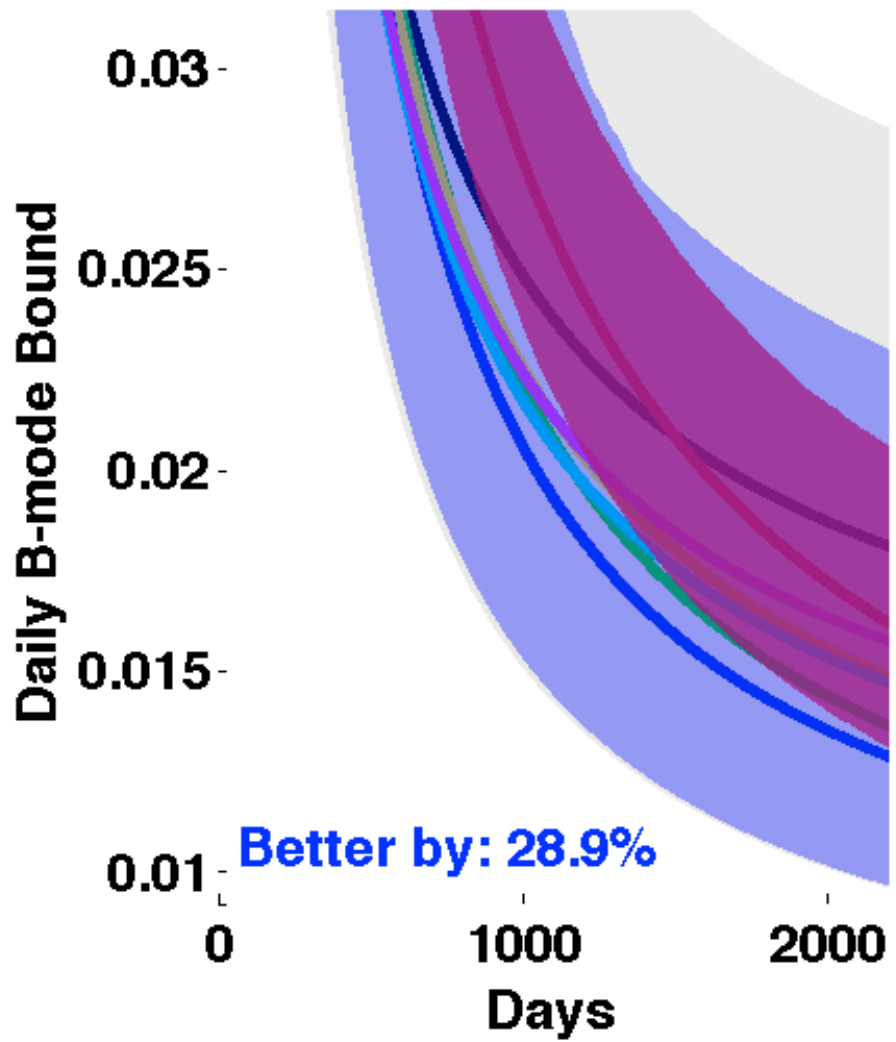}
\caption{The average smallest detectable tensor-to-scalar
     ratio with different adaptive survey (bandit) strategies.  The greedy method is
     drawn in black, with its best and worst performances
     bracketing the area shaded in gray. Greedy with
     $\epsilon=1$ is drawn in red and its performance interval
     is also shaded in red. The best-performing method in all
     scenarios is UCB and its performance range is shaded in blue.
     We also note the best  improvement achieved with respect to greedy. 
     {\it Column 1} is for the worst-case $10\%$ normalization scenario, 
     {\it Column 2} is for the conservative scenario of $3.6\%$
     normalization and {\it Column 3} is for the optimistic case
     of $3.6\%$ normalization and $80\%$ de-lensing
     ($\alpha=0.2$ in Eq.~(\ref{eq:r})). {\it Top Row}:
     Experiment 1, with $n_p=10$ patches and step-size of three
     observation days. {\it Middle Row}: Experiment 2, with
     $n_p=15$ patches and step-size of six observation
     days. {\it Bottom Row}: Experiment 3, $n_p=5$ and the
     step-size is three days.}
\label{fig:figureofmerit}
\end{figure*}

As we explained above, our overall figure-of-merit in comparing adaptive survey 
strategies is the smallest detectable tensor-to-scalar
ratio $r$ at ``1-sigma'' confidence. In
Fig.~\ref{fig:figureofmerit}, we plot our results for the three
experiments in Table~\ref{table:exppars}, for three scenarios: a
pessimistic scenario with $10\%$ dust polarization fraction and
no de-lensing, a conservative scenario with $3.6\%$
normalization and no de-lensing and an optimistic scenario with
$3.6\%$ normalization and a $20\%$ lensing residual. In all cases, 
the UCB method achieves the best results
for all experiments (though it is very close to the Boltzmann
method in many instances). In the case of experiment 2 in the conservative scenario, even the best-case performance of the greedy method is inferior to the averages of the UCB method (and also $\epsilon$-greedy with $\epsilon=1$). This is because with this low-noise experiment, prolonging the integration, even over the cleanest patch, only goes so far and it is often preferable to spend equal time on the two (or more) cleanest patches.
   
We see that the improvement on average when
comparing to the greedy method ranges from $\sim25\%$ to $\sim70\%$
(and it goes without saying that it will only improve further if the 
observation is prolonged). Thus, adaptive  strategies have 
great potential to improve the sensitivity in IGW searches.

\section{Discussion}

\subsection{Simplifying assumptions and caveats}

It is worthwhile to discuss several simplifications we have made throughout the analysis above. First, we have made exclusive use of the FGPol templates \cite{O'Dea:2011kx} both to motivate the need for adaptive survey strategies and in order to estimate the expected distribution of foreground levels across the sky. The use of other templates that have been developed in the literature \cite{Schlegel:1997yv,Fantaye:2011zq,Delabrouille:2012ye}, or the recent high-frequency polarization measurements by the Planck experiment (which are not sensitive enough to efficiently constrain the polarization dust amplitudes in low intensity regions), may lead to somewhat different results, but most likely will not alter our conclusions. In the future, as better data is accumulated, it will be worthwhile to consider improvements upon the methods presented above. For example, one way to incorporate prior data is in the initialization of the different action-value estimates of the targeted patches, which will ensure faster convergence onto optimal patches.

Secondly, we have limited our analysis to the case of experiments operating at a single frequency. This allowed us to focus on a single source of foregrounds and work under the assumption that it could not be efficiently removed through component separation or other methods. Most B-mode experiments will eventually gather measurements from multiple frequencies in order to enable foreground subtraction. In this case, the foreground amplitudes we have considered here will be replaced by the amplitude of the remaining residuals  \cite{KovKamExploration}. In a very optimistic scenario, where these would be smaller than the IGW B-mode signal, the advantage of our methods would be restricted. In \cite{KovKamExploration}, we elaborate on this issue and present a different approach, whereby an independent stage of exploration at higher-frequencies is performed \textit{before} the stage of prolonged observation. 

Additional assumptions include the neglect of E and B mode mixing, which may pose a problem in partial sky measurements \cite{Amblard:2006ef,Bunn:2002rx,Zhao:2010sza} (although optimized estimators greatly reduce this complication \cite{Smith:2005gi}) and the approximation that the likelihood curves for the power spectra estimators are Gaussian (in our B-mode simulations we simply drew a measured amplitude from a normalized distribution matching the likelihood mean and width, as opposed to drawing the estimated value for each multipole from a corresponding $\chi^2$ distribution). We do not expect these assumptions to have a meaningful impact on our results. 

Finally, we have limited our analysis to just a handful of scenarios, ranging from pessimistic to optimistic, in an attempt to provide a complete picture regarding the prospects of the methods proposed. We have verified that changing some of the parameters chosen here, such as the power-law index for the PED power spectrum, the level of dust polarization fraction outside the Galactic plane and the efficiency of the de-lensing process, do not substantially affect our conclusions. However, a more detailed study should be conducted when implementing these methods to a specific B-mode experiment. One enhancement upon our analysis is to calculate the full foreground power spectrum directly from the template patch, instead of relying on a best-fit power law. Another important ingredient to include is the cost of moving between different sky patches when calculating the total regret of different observing strategies. While this may degrade the improvement in sensitivity somewhat, it will be offset by the continued gain from identifying the optimal patches for observation as the experiments progress to their next stages, which we have not taken into account here. 
 
 \subsection{Adaptive strategies elsewhere}

Analogous adaptive survey strategies might also be used to help
detect 21-cm intensity fluctuations from the EoR
\cite{Furlanetto,Morales}, where the signal must
be distinguished, using angular and/or frequency information, 
from a variety of Galactic and extragalactic foregrounds
\cite{Jelic:2008jg,Dimatteo:2001gg,Oh:2003jy,Santos:2004ju,Bowman:2008mk,Liu:2011hh,Gleser:2007fg}
that are expected to be larger by several orders of magnitude.
The desired signal is expected to extend from several
arcminutes to several degrees.  What is currently envisioned,
for example, for the Low Frequency Array \cite{LOFAR} and the Square Kilometer Array \cite{SKA}, is a deep integration on a
$\sim5^\circ\times 5^\circ$ region of the sky.
As is well known by now, the amplitude of Galactic foregrounds may vary
considerably from one such region on the sky to another.  While
existing Galactic dust and synchrotron maps can be used to
identify regions of the sky that are likely to be clean, again,
detailed maps at the several-hundred MHz frequencies of
interest, with the angular resolutions of interest, do not
exist.  There are also likely to be foregrounds in frequency
space contributed by radio recombination lines \cite{Oh:2003jy},
and little is known about their distribution at the high
Galactic latitudes that present the likeliest targets for these
deep EoR integrations.  The distribution of the
extragalactic-foreground intensity across the sky is expected to
be smoother than the Galactic foregrounds.  Even so, clustering
of highly-biased radio sources at high redshift may give rise to
significant variation in foreground intensities from one patch
of the sky to another.  The aim of an adaptive bandit strategy would in
this case be to identify the cleanest patch in a fixed survey time\footnote{Recently, the performance of two non-adaptive modes of observation for the Murchinson Widefield Array \cite{MWA} --- corresponding to the greedy and $\epsilon$-greedy with $\epsilon=1$ methods described above --- was investigated in detail \cite{Thyagarajan:2013eka}.}.  

Another example of a foreground-limited stochastic measurement is 
that of a gravitational wave background (GWB) \cite{PrimGWBack1,PrimGWBack2}, which may be pursued by terrestrial experiments such as The Laser Interferometer Gravitational Wave Observatory \cite{LIGO} and several Pulsar Timing Array consortia \cite{IPTA}. In the relevant frequency bands, the stochastic GWB signal may be obscured by confusion foreground stemming from bright sources, including supermassive black hole binaries \cite{Jenet:2006sv}, neutron star binaries \cite{Regimbau:2005ey} and Galactic white dwarf binaries \cite{Evans}. In an effort to minimize the effect of foregrounds, adaptive strategies may be used to improve the sensitivity of the relevant ground-based experiments.

Likewise, similar strategies should be explored for a JWST deep
field aimed to detect the first stars/galaxies
\cite{Bromm:2003vv,Barkana:2000fd}.  CMB measurements of the
reionization optical depth suggest that the first stars formed at
redshfits $z\sim 10$, and these stars should emit radiation
that today falls in the wavelengths probed by JWST
\cite{Santos:2002hd,Trenti:2009cj}.  Still, the abundance of
these objects, their luminosity function, etc., are largely a
matter of guesswork.  It may well be that the population of
the first stars/galaxies will fill the entire image, or that there may
be a vast population of low-flux sources from lower redshifts
that may obscure the EoR objects of interest, as suggested in
Ref.~\cite{Windhorst:2007}.  The very faint EoR sources, in
this case, will be easiest identified in the cleanest field on
the sky.  Again, an adaptive survey strategy to identify the
cleanest of several candidate patches may well be warranted.

In this regard, one should be mindful of an inherent difference between the application of adaptive survey strategies to stochastic measurements and deep-field imaging. While a choice of an optimal sky patch for deep-field imaging may considerably improve the sensitivity of a given experiment, the initial stage of exploration is mostly wasted. Therefore, the overall figure-of-merit when comparing different strategies should be the speed of convergence onto the optimal patch instead of the total regret, and the best-performing algorithms may be different. Nevertheless, an initial stage of exploration may still prove invaluable in reducing the error in the integrated signal from the chosen patch to image. 

A somewhat different kind of MAB-like problem in the context of astronomical observations shows up in the case of follow-up observations of transients \cite{Djorgovski:2011rv}, for example. In this case, the exploration versus exploitation tradeoff is manifested in the challenge of allocating the wide array of resources to the task, each with very different characteristics and cost functions. Hence, the overall figure-of-merit for comparing various algorithms may once again be different than considered here. 

We leave the detailed study of the prospects of adaptive strategies in these setups to future work.

\section{Conclusion}

The era of CMB B-mode measurements has only just begun \cite{Zaldarriaga:1998ar,SPTpolBModes, Ade:2014afa,Ade:2014xna}. The race for the detection of IGW B modes is picking up and the competing efforts will be mainly characterized by their ability to tighten the upper bound on the tensor-to-scalar ratio. This endeavor is extremely difficult and will grow more so as the upper bound is decreased, particularly due to the major obstacles which need to be efficiently removed --- foreground contamination and lensed E modes --- whose amplitudes are independent of the targeted IGW B-modes. Therefore, any novelty in the approach to conducting these measurements which may mitigate these problems is both timely and important. 

In this paper we have proposed exactly such a novelty, in the form of adaptive survey (bandit) strategies, inspired by heuristic solutions to the MAB problem. Focusing on polarized foreground components, whose amplitudes are predicted to vary considerably across the sky, our target was to balance the time spent on exploration to find lower-noise patches of sky with that devoted to exploitation of the optimal patches through deep integration. 
By formulating this tradeoff as a machine-learning problem, we were able to adopt heuristic algorithms developed in the general context of the MAB challenge and implement them in the settings of fiducial B-mode experiments.

Attempting to go beyond a mere proof-of-concept, we have incorporated a number of crucial ingredients into our analysis, such as prior knowledge from existing surveys, several fiducial sets of instrumental properties, different possible forecasts regarding the average amplitude of foregrounds, and realistic prospects for the level of lensing residuals (including pessimistic cases of no de-lensing). 
Relying on advanced templates for polarized emission from dust in the Galaxy, we demonstrated that single frequency experiments could improve their upper bounds on the tensor-to-scalar ratio by factors of a few on average and even higher when comparing worst-case performances of standard methods versus the ones proposed here. 

The assumptions made in this work were discussed in the previous section, along with some possible caveats in the implementation of the adaptive survey strategies proposed here in B-mode experiments. While the precise improvement these methods may enable remains to be seen, we feel that the case has been made for their consideration in any ground-based experimental setup focusing on CMB B-mode detection with partial sky coverage.
 
Lastly, as the tradeoff between exploration and exploitation shows up in other realms of observational cosmology as well, this work opens the door for additional applications of adaptive survey strategies. These may include other stochastic measurements, such as the power spectrum of 21-cm fluctuations \cite{Furlanetto, Morales, Pritchard} or a primordial GWB \cite{PrimGWBack1,PrimGWBack2}, deep-field imaging by radio interferometers \cite{LOFAR,SKA} or optical telescopes \cite{Windhorst:2005as,Stiavelli}, and the allocation of limited resources in follow-ups of identified astrophysical transients \cite{Djorgovski:2011rv}, to name a few.

\appendix

\section{B Mode Measurements}

\subsection{Instrumental Noise}

The instrumental noise in a CMB-polarization experiment is
determined by the detector-array sensitivity (or
noise-equivalent temperature NET) $s$, its angular
resolution $\theta_{\rm fwhm}$, the sky coverage $f_{\rm sky}$
and the total observation time $T$ (which is reduced in practice
by the observing efficiency).  
The pixel noise $\sigma_{\rm pix}=s/\sqrt{t_{\rm pix}}$ is determined by the detector sensitivity $s$ and the observation time $t_{\rm pix}=T/N_{\rm pix}$ dedicated to each pixel. Defining the inverse weight per solid angle, $w^{-1}(T)=4\pi
s^2/T$, the angular power spectrum of the instrumental noise, assuming the experimental beam is approximately Gaussian in shape, is given by \cite{Tegmark:1997vs}
\be
C^N_{\ell}=\frac{\Omega\sigma_{\rm pix}^2}{N_{\rm pix}}e^{\ell^2\sigma_b^2}=\frac{\Omega s^2}{T}e^{\ell^2\sigma_b^2}=f_{\rm sky}w^{-1}(T)e^{\ell^2\sigma_b^2},
\ee
where $\Omega=4\pi f_{\rm sky}$ and $\sigma^2_b=\theta^2_{\rm fwhm}/(8\ln{2})$.

\subsection{Statistical Estimators on Partial Sky}

The measurements we deal with are those of power spectra. In a
maximum-likelihood analysis, the Fisher forecast for the error
in the measurement of the amplitude $A$ of a power spectrum
$C_\ell$ is given by \cite{Jaffe:2000yt},
\be\label{eq:smallamp}
     \frac{1}{\sigma_A^2} = \sum_\ell \left( \frac{ \partial C_\ell }{ \partial A} \right)^2 \frac{1}{\sigma_\ell^2}.
\ee
We will work under the assumption that the likelihood function
is Gaussian in the vicinity of its maximum
\cite{Jungman:1995bz,Zhao:2008re}. To emphasize this
approximation we use quotation marks around ``1-sigma'' when
referring to the error $\sigma_A$.  

In order to choose between exploration and exploitation on the
fly, we need to estimate the PED amplitude $A$ in a targeted
patch within the observation time allotted to a {\it single step} of the experiment. 
For a given sky coverage $f_{\rm sky}$, the ``1-sigma'' error for an individual 
$\ell$ in the estimated value $\widehat{A}$ is \cite{Jaffe:2000yt,Knox:2002pe,Kesden:2002ku} 
\be
  \sigma_\ell^{\widehat{A}} = \sqrt{\frac{2}{ f_{\rm
  sky}(2\ell+1)}}\left(\alpha C^L_\ell +
  f_{\rm sky} w^{-1}(t_{\rm step}) e^{\ell^2\sigma_b^2}\right),
\ee
where $C^L_\ell$ is the lensing B-mode
contribution and $w^{-1}(t_{\rm step})$ is the inverse weight per solid angle given $t_{\rm step}$ observation time. 
The quantity $1-\alpha$ parametrizes the level of
de-lensing \cite{Sigurdson:2005cp,Smith:2010gu} that was applied to the
data. When comparing forecasts for upper bounds on the
tensor-to-scalar ratio, we will consider $\alpha=0.2$ and
$\alpha=1$ as best and worst-case scenarios, respectively.  The
total ``1-sigma'' error in the measurement of $\widehat{A}$
over a time $t_{\rm step}$ is thus,
\be
     \sigma^{\widehat{A}}=\left[\frac{f_{\rm sky}}{2}
     \sum\limits_{\ell_{\rm min}}^{\ell_{\rm max}}
     \frac{(2\ell+1)(\tilde{C}^D_\ell)^2}{\left(\alpha
     C^L_\ell+f_{\rm sky}w^{-1}(t_{\rm step})e^{\ell^2
     \sigma_b^2}\right)^2}\right]^{-\frac{1}{2}},
\label{eq:estimatorerror}
\ee
where $\tilde{C}_{\ell}^D=C^D_\ell/A =
2\pi\ell^{-m}/[\ell(\ell+1)]$ encodes the $\ell$ dependence of the
PED power spectrum and $\ell_{\rm min}=180/\theta$ is the largest scale accessible
by an experiment with sky coverage $f_{\rm sky}=\theta^2$.

\begin{figure}[h!]
\includegraphics[width=0.49\linewidth]{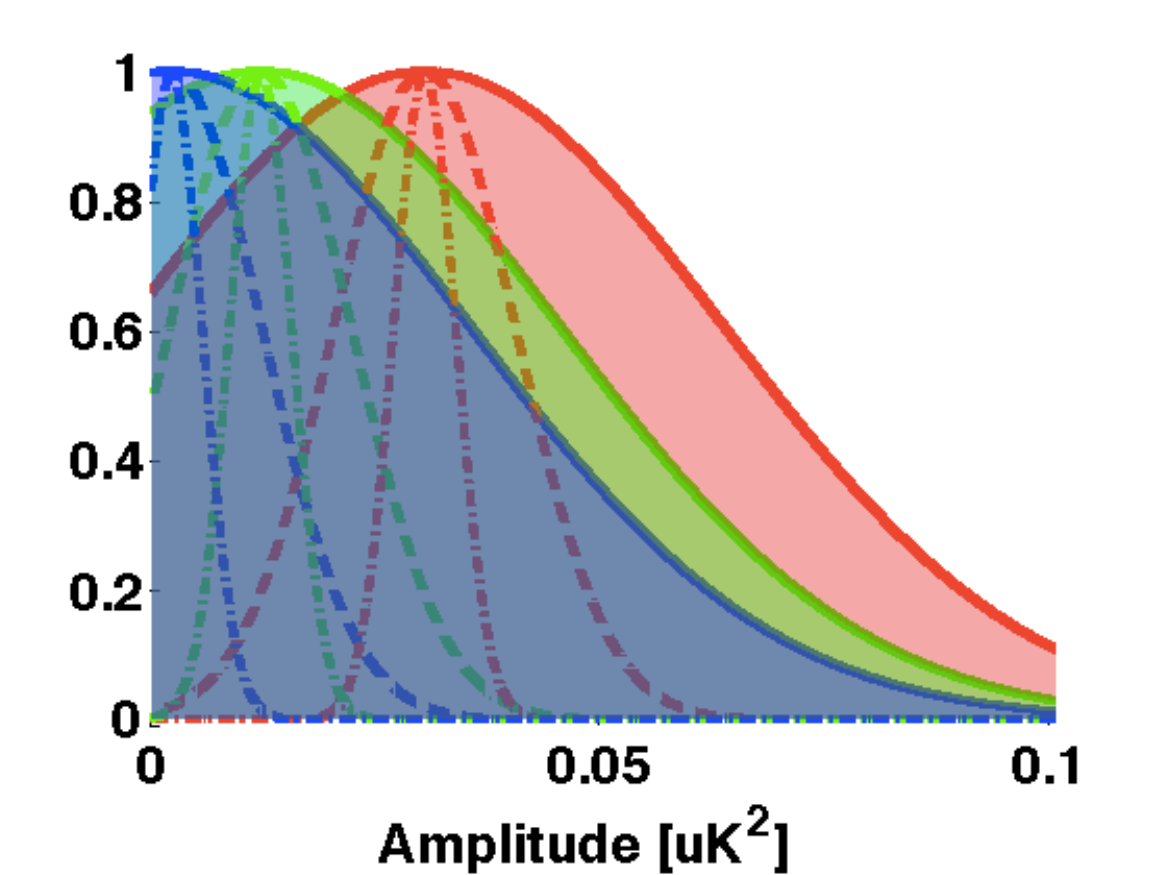}
\includegraphics[width=0.49\linewidth]{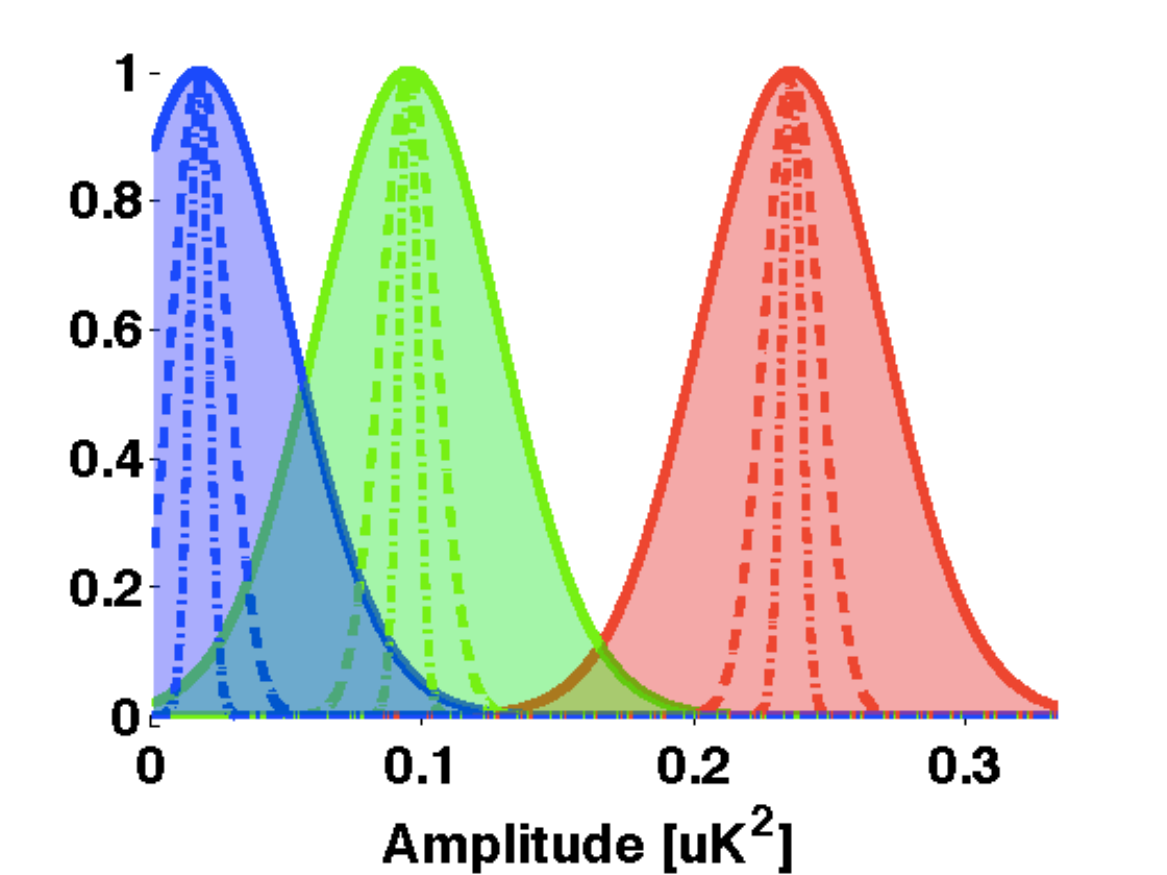}
\caption{{\it Left}: Likelihood functions for the measured amplitude values for the maximum-, median-, and minimum-amplitude patches (in blue, red and green) from our sample of $19$ non-overlapping $15^{\circ}\times15^{\circ}$ PED patches in the conservative scenario of $3.6\%$ average dust polarization fraction, given the instrumental noise of Experiment 1. Widths correspond to step sizes of 3, 10 and 30 days of observation (in solid, dashed and dot-dashed, respectively). {\it Right}: Same plots for the worst-case $10\%$ average dust polarization fraction, in which case it is much easier to discern between the patches.}
\label{fig:likelihoods}
\end{figure}

In Fig.~\ref{fig:likelihoods} we plot the normalized (Gaussian)
likelihood curves for the measured amplitudes of the patches with maximum, median, and minimum PED amplitudes in the FGPol template (see Fig.~\ref{fig:SignalAndNoise}). The goal of the adaptive survey (bandit) 
strategies will be to distinguish between the means of these distributions 
in order to spend more time observing lower-noise patches. Clearly, this
task is harder when the similarity between the distributions is
larger (we will quantify this in the next Section).

Our figure of merit when comparing the prospects of different
adaptive survey strategies will be the smallest amplitude of
primordial B modes that can be distinguished from the null
hypothesis at ``1-sigma'' confidence after a fixed total
observation time $T$. Denoting the $\ell$ dependence of the IGW B-mode
power spectrum by $\tilde{C}^B_\ell$, the smallest amplitude in
a patch $p$ detectable at ``1-sigma'' (with a total $t_p$ observation time spent on that patch), according to
Eq.~(\ref{eq:smallamp}), is then
\be
     \sigma_p^r=\left[\frac{f_{\rm sky}}{2}
     \sum\limits_{\ell_{\rm min}}^{\ell_{\rm
     max}}\left(\frac{\sqrt{(2\ell+1)}\tilde{C}^B_\ell}{A_p\tilde{C}^D_\ell+\alpha
     C^L_\ell+f_{\rm
     sky}w(t_p)^{-1}e^{\ell^2\sigma_b^2}}\right)^2\right]^{-\frac{1}{2}}
\label{eqn:sigmaprone}
\ee
Notice that we set the sample variance of the primordial signal
to zero in this expression (since we are comparing to the null
hypothesis).\footnote{Note that in a realistic experiment, the
dust amplitudes $A_p$ must be determined from the data in tandem
with the amplitude $r$, and the minimal upper bound that can be placed on $r$ is thus degraded
relative to that suggested by
Eqs.~(\protect\ref{eqn:sigmaprone}) and (\protect\ref{eq:r}).
Still, Eqs.~(\protect\ref{eqn:sigmaprone}) and
(\protect\ref{eq:r}) can provide a useful overall figure-of-merit for
comparing roughly the {\it relative} sensitivities of different
experiments (as done in \cite{Wu:2014hta}), and of different observational strategies.}

Finally, if an adaptive survey ends up spending a different observing
time $t_p$ (where $\sum\limits_{p=1}^{n_p}t_p=T$) on each of
$n_p$ patches with known dust amplitudes
$A_p={A_1,\dots,A_{n_p}}$, then the smallest detectable
primordial B-mode amplitude will be
\be
\sigma^{\rm r}=\left[\frac{f_{\rm sky}}{2} \sum\limits_{p=1}^{n_p} \sum\limits_{\ell_{\rm min}}^{\ell_{\rm max}}\left(\frac{\sqrt{(2\ell+1)}\tilde{C}^B_\ell}{A_p\tilde{C}^D_\ell+\alpha C^L_\ell+f_{\rm sky}w(t_p)^{-1}e^{\ell^2\sigma_b^2}}\right)^{2}\right]^{-\frac{1}{2}}
\label{eq:r}
\ee
As we clarify above, this expression for $\sigma^{\rm r}$ is
{\it never} used during the experiment by the algorithm employed to determine the allocation of
time to different patches.  We calculate it {\it only after the experiment has ended and the total observation time
has been allocated by the different adaptive strategies} as an overall
figure-of-merit to estimate how well the strategies have done in
improving the sensitivity to $r$.  As we furthermore clarify
below, the adaptive (bandit) algorithms never assume any prior knowledge of
the $A_p$s.  All information with which decisions are made about
time allocations come from the measurements themselves.


\end{document}